\documentclass[useAMS,usenatbib]{mn2e}
\usepackage{amsmath}
\usepackage{graphicx}
\usepackage{epstopdf}
\usepackage{amssymb,latexsym}

\title[Binary fractions in the massive young star clusters NGC 1805
  and NGC 1818]{The binary fractions in the massive young Large
  Magellanic Cloud star clusters NGC 1805 and NGC 1818}

\author[Chengyuan Li, Richard de Grijs and Licai Deng]{Chengyuan
  Li$^{1,2,3}$\thanks{E-mail: joshuali@pku.edu.cn}, Richard de
  Grijs$^{1,2}$ and Licai Deng$^{3}$\\ 
$^{1}$Kavli Institute for Astronomy and Astrophysics, Peking
  University, Yi He Yuan Lu 5, Hai Dian District, Beijing 100871,
  China\\ 
$^{2}$Department of Astronomy, Peking University, Yi He Yuan Lu 5, Hai
  Dian District, Beijing 100871, China\\ 
$^{3}$Key Laboratory for Optical Astronomy, National Astronomical
  Observatories, Chinese Academy of Sciences, 20A Datun
  Road,\\ Chaoyang District, Beijing 100012, China}
\begin{document}


\pagerange{\pageref{firstpage}--\pageref{lastpage}} \pubyear{2013}

\maketitle

\label{firstpage}

\begin{abstract}
Using high-resolution data sets obtained with the {\sl Hubble Space
  Telescope}, we investigate the radial distributions of the F-type
main-sequence binary fractions in the massive young Large Magellanic
Cloud star clusters NGC 1805 and NGC 1818.  We apply both an
isochrone-fitting approach and $\chi^2$ minimization using Monte Carlo
simulations, for different mass-ratio cut-offs, $q$, and present a
detailed comparison of the methods' performance.  Both methods yield
the same radial binary fraction profile for the same cluster, which
therefore supports the  robustness and applicability of either
method to young star clusters which are as yet unaffected by the
presence of multiple stellar populations. The binary fractions in
these two clusters are characterized by opposite trends in their
radial profiles. NGC 1805 exhibits a decreasing trend with increasing
radius in the central region, followed by a slow increase to the
field's binary-fraction level, while NGC 1818 shows a monotonically
increasing trend. This may indicate dominance of a more complicated
physical mechanism in the cluster's central region than expected a
priori. Time-scale arguments imply that  early dynamical mass
segregation should be very efficient and, hence, likely dominates the
dynamical processes in the core of NGC 1805. Meanwhile, in NGC 1818
the behavior in the core is probably dominated by disruption of soft
binary systems. We speculate that this may be owing to the higher
velocity dispersion in the NGC 1818 core, which creates an environment
in which the efficiency of binary disruption is high compared with
that in the NGC 1805 core.
\end{abstract}


\section{Introduction}

Characterization of the binary fractions in star clusters is of
fundamental importance for many fields in astrophysics. Observations
indicate that the majority of stars are found in binary systems
\citep{Duqu91,Grif03,Halb03,Kouw05,Rast10}, while it is generally
accepted that most stars with masses greater than 0.5 M$_{\odot}$ are
formed in star clusters \citep{Lada03}. In addition, since binaries
are on average more massive than single stars, in resolved star
clusters these systems are thought to be good tracers of (dynamical)
mass segregation. Over time, dynamical evolution through two-body
relaxation will cause the most massive objects to migrate to the
cluster centre, while the relatively lower-mass objects remain in or
migrate to orbits at greater radii
\citep{grijs02a,grijs02b,grijs02c}. This process will globally
dominate a cluster's stellar distribution. However, close encounters
involving binary systems may disrupt `soft' (i.e., generally wide)
binaries \citep{Hegg75, Ivan05, Tren07b, Park09, Kacz11,
  grijs13}. This process will occur more frequently in a cluster's
central, dense region than in its periphery, which may mask the
effects of mass segregation. In the possible presence of
intermediate-mass black holes in cluster cores, this dynamical
scenario will proceed even more rapidly \citep{Tren07c}.

Young massive clusters are good targets to explore the early
observational signatures of both (early) dynamical mass
segregation and binary disruption, particularly since old clusters
have already experienced significant evolution and their member stars
will likely have reached a dynamical state close to energy
equipartition. Although the Milky Way hosts a small number of young
massive ($\gtrsim 10^5$ M$_\odot$) clusters, such as Westerlund 1,
these objects are not suitable to explore these issues in detail. The
rather extreme reddening associated with Westerlund 1 and other young,
intermediate-mass clusters near the Galactic Centre prevents us from
sampling significant numbers of their member stars in sufficient
detail. This makes the young clusters in the Large Magellanic Cloud
(LMC) interesting in this context. However, because the components of
binary systems in massive, crowded star clusters at the distance of
the LMC cannot be resolved with current instruments, binary fractions
in distant, young massive clusters have not yet been studied as
thoroughly as those in their older Galactic counterparts.

A number of approaches have been developed to analyse binary fractions
in clusters. One of these aims at identifying binaries by measuring
their radial velocity variability \citep{Mateo96}, but this method is
only sensitive to bright stars because of spectroscopic limits, and it
is affected by a strong detection bias towards short-period
binaries. Another approach explores the photometric variability of
cluster members. However, this method is also biased towards binaries
with short periods \citep{Milo12} and the relatively high orbital
inclinations required to observe at least partial eclipses, which
cause variability. Both approaches are hampered by low discovery
efficiencies, since they require extensive time-domain observations
and, hence, significant allocations of telescope time.

A promising alternative method is based on statistical analysis of
cluster colour--magnitude diagrams (CMDs). Unresolved binaries will be
found on the brighter and redder side of the single-star main sequence
(MS). Their importance can be quantified through careful Monte
Carlo-type analysis. This approach can potentially yield a high
detection efficiency, since it only depends on a single set of
observations. In principle, one does not need to adopt many physical
assumptions either; the approach is only based on our understanding of
stellar evolution. This also allows us to simulate artificial
observations for comparison, without having to rely on too many
detailed physical considerations, such as regarding the period
distribution of the binaries or their orbital inclinations
\citep{Rube97, Zhao05, Soll07, Soll10, Milo12}. However, almost all
clusters which have been analysed based on the latter method are old
stellar systems, in which dynamical evolution is expected to have
significantly altered the initial binary population. Efforts have
recently begun to address this issue in the context of the much more
distant, young massive clusters in the LMC
\citep{Elso98,Hu10,Hu11,grijs13}.

In this paper, we analyse the binary-fraction properties of two young
massive LMC clusters, NGC 1805 and NGC 1818. Previous studies of these
clusters have reported that both are significantly mass-segregated
\citep[e.g.,][]{grijs02a,grijs02b,grijs02c}. \cite{Elso98},
\cite{Hu10} and \cite{grijs13} studied the properties of the binary
systems in NGC 1818; although \cite{Elso98} reported apparent evidence
of `binary segregation' of the most massive binary systems in the
cluster, \cite{grijs13} found the opposite effect. In this paper, we
explore the binary properties in both clusters in much more detail,
while for NGC 1818 in particular we reduce the previously reported
uncertainty ranges significantly. We first analyse their CMDs to
estimate the ratio of the number of possible binaries to the total
stellar sample in the same magnitude range, simply based on assessment
of the best-fitting isochrones and the expected photometric colour
spread. We have also developed a more sophisticated and more accurate
approach based on $\chi^2$ minimization using Monte Carlo
simulations. We will show that both methods lead to the same result
for the same cluster, although the $\chi^2$ minimization is preferred
because it is physically better justified.

This paper is organized as follows. The data reduction is discussed in
Section 2. The details of the isochrone-fitting and
$\chi^2$-minimization methods are described in Sections 3 and 4,
respectively. We also discuss and show how well either method
performs. We go well beyond the approaches taken in previous studies
of our target clusters. We have updated and significantly improved our
analysis method with respect to that employed in \cite{grijs13}. We
also take the possible effects of fast stellar rotation into
consideration and show that the results remain robust, even if rapidly
rotating stars were to contribute significantly to the clusters'
CMDs. We discuss the detailed physical implications of our results, as
well as any possible shortcomings of the methods applied, in Section 5
and conclude the paper in Section 6.

\section{Data Reduction}

The data sets pertaining to NGC 1805 and NGC 1818 were obtained from
{\sl Hubble Space Telescope (HST)} programme GO-7307 (PI Gilmore). The
complete data set for a given cluster is composed of a combination of
three Wide Field and Planetary Camera-2 (WFPC2) images in the F555W
and F814W filters, which roughly correspond to the Johnson--Cousins
$V$ and $I$ bands, respectively, with total exposure times of 2935 s
and 3460 s. Both data sets consist of long- and short-exposure
pointings in which the Planetary Camera (PC) chip was centred on the
cluster centre region (referred to as our `CEN' fields), with a third,
longer-exposure image centred on the cluster's half-light radius on
one side of the cluster centre \citep[`HALF'; for more details,
  see][]{grijs02a}. Images with a total exposure time of 1200 s ($V$)
and 800 s ($I$) of a nearby field region centred on the LMC's stellar
disc were also used to assess and correct for background contamination
\citep[cf.][]{Hu10}. Figures 1 and 2 show the spatial distributions of
the stars associated with the NGC 1805 and NGC 1818 cluster fields,
respectively.

\begin{figure}
 \includegraphics[width=95mm]{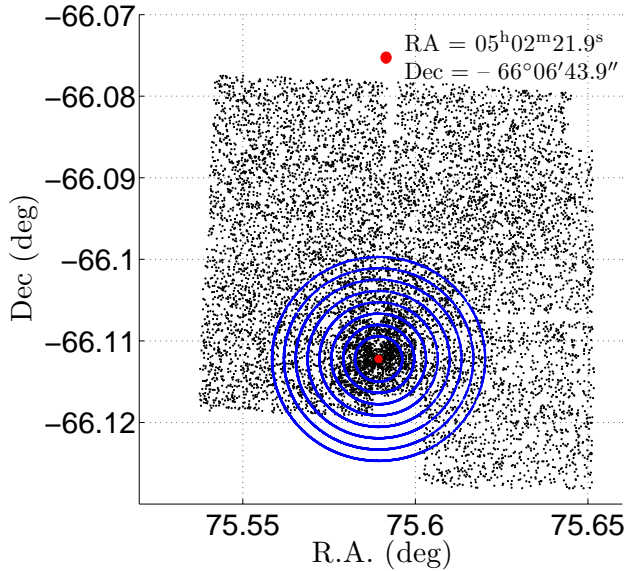}
 \caption{Spatial distribution of the stars in the NGC 1805 cluster
   field. Blue circles are drawn at radii from 10 to 45 arcsec in
   steps of 5 arcsec. The solid red bullet represents the cluster
   centre.}
  \label{F1.eps}
\end{figure}

\begin{figure}
 \includegraphics[width=95mm]{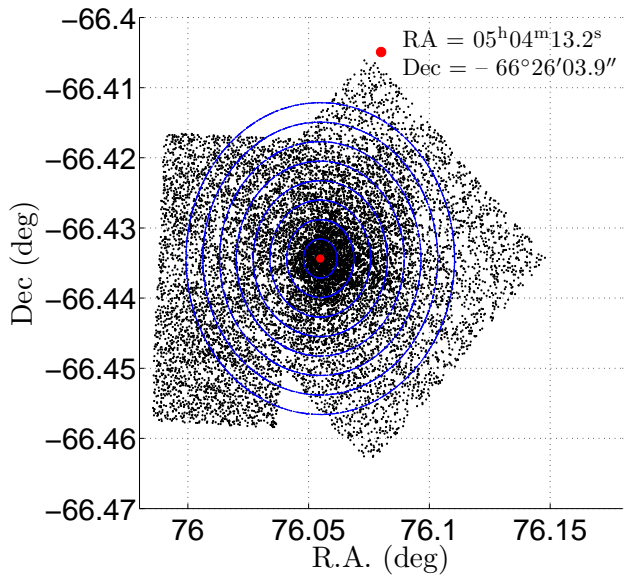}
 \caption{As Fig. 1, but for NGC 1818. The blue circles range from 10
   to 80 arcsec in steps of 10 arcsec.}
  \label{F2.eps}
\end{figure}

Photometry was performed with {\sc HSTphot} \citep{Dolp00}, a
specialized photometry package for analysis of {\sl HST}/WFPC2 data
sets \citep[for details, see][]{Dolp05,Hu10}, which can automatically
deliver robust {\sl HST}/WFPC2 photometry as well as the corresponding
photometric uncertainties. For each cluster, we obtained point-spread
function photometry for the three raw images, which thus resulted in
three distinct stellar catalogues. Since bright stars are saturated in
the long exposures and faint stars cannot be detected in the
short-exposure images, we combined the three catalogues into a
complete master data set, retaining only stars with photometric
uncertainties up to 0.22 mag. If a given star appeared more than once
in our catalogues, we adopted its magnitude and photometric
uncertainty based on the deeper observation, which was usually
associated with smaller photometric uncertainties. We verified that,
in such cases, both entries in our catalogues referred to the same
star based on both their magnitudes and spatial coordinates. The
spatial distributions of the stars in both cluster fields are shown in
Figs 1 and 2.

\begin{table}
 \centering
 \begin{minipage}{85mm}
  \caption{Stellar numbers in the NGC 1805 and NGC 1818 catalogues.}
  \begin{tabular}{@{}lcc@{}}
  \hline
   Name     &  NGC 1805 & NGC 1818 \\
 \hline
 CEN short & 1523 & 2264\\
 CEN long & 5785 &  7189\\
 HALF short & 6770 & 8755\\
 Nearby field & 4115 & 5268\\
 Combined & 10770 & 14067 \\
Decontaminated & 6849 & 9014 \\
\hline
\end{tabular}
\end{minipage}
\end{table}

The photometry of the nearby background field was processed similarly
\citep[for details about the background fields, see][]{Cast01}. We
decontaminated the cluster CMDs using a method similar to that
employed by \cite{Kerb05} and \cite{Hu10}. We divided both the cluster
and field CMDs into a carefully considered number of cells and counted
the number of stars in each. For our target clusters, the colour range
for both the field and the cluster CMD is approximately 3 mag, while
the cluster CMD's full magnitude range covers more than 10 mag. We
divided the cluster and field CMDs into 50 cells in colour and 100
cells in magnitude, with cell sizes of 0.1 and 0.15 mag,
respectively. We adopted these cell sizes because of the comparable
colour difference between the single-star and binary MSs (see Section
3.2). Small changes in cell size will not affect our decontamination
efficiency. However, if we were to define significantly larger cells,
this would prevent us from distinguishing substructure in the CMD
caused by the binary population, while a significant reduction in cell
size would increase stochastic effects and, hence, introduce
random-number fluctuations in our background decontamination
procedure.

After correcting for the difference in areal coverage, we randomly
removed a number of stars corresponding to that in the area-corrected
field-star CMD from the cluster CMD. In some bins, the number of field
stars was larger than the equivalent number of cluster stars. In such
cases, we removed all cluster stars. Although this may lead to
increased uncertainties, this situation only occurred near the edges
of the relevant area in colour--magnitude space explored; in the
region of interest in this paper, covering the genuine single-star and
binary MSs, the numbers of cluster stars always exceed those of the
field stars. Although the exposure times for the cluster and
background CMDs are not identical, here we only investigate the binary
properties in a small magnitude range (see below), where both data
sets are complete. In Fig. 3 (top row) we display, as an example, the
raw CMD of NGC 1805 (left), the CMD of the nearby field region
(middle) and the decontaminated CMD (right); for NGC 1818, see fig. 3
of \cite{Hu10} for an equivalent presentation. We summarize the number
of stars in the different raw catalogues and those in the combined
catalogues considered for both clusters in Table 1.\footnote{Since the
  magnitude range of the background catalogue is not identical to that
  of the combined catalogue, the decontaminated catalogue contains
  more stars than the difference between the background and cluster
  catalogues. For the analysis performed in this paper, we use the
  decontaminated, combined catalogue.} The panels in the bottom row of
Fig. 3 show enlargements of the full CMDs (top row), focussing on the
magnitude range of interest for our binarity analysis. In the latter
panels, we have also indicated the cell sizes used for our background
decontamination.

\begin{figure}
 \includegraphics[width=95mm]{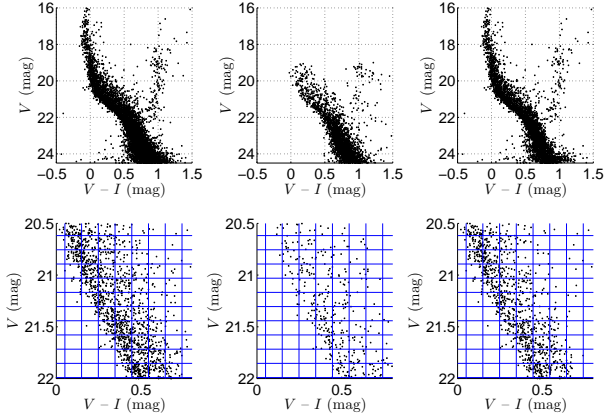}
 \caption{(top left) Raw CMD of NGC 1805. (top middle) CMD of the
   nearby field region. (top right) Decontaminated CMD of NGC
   1805. (bottom) Same as the top panels, but highlighting the
   magnitude range $V \in [20.5, 22.0]$ mag. We have also delineated
   the cell sizes used for our background decontamination.}
  \label{F3.eps}
\end{figure}

\section{Isochrone Fitting}

\subsection{Basic considerations}

Distant binaries in crowded environments, such as in compact LMC
clusters, cannot be resolved into their individual components with
current instrumentation. The fluxes of the individual binary
components are thus returned as a single photometric measurement for a
given binary system,
\begin{equation}
  m_{\rm b}=-2.5 \log(10^{-0.4m_1}+10^{-0.4m_2}), 
\end{equation}
where $m_1$ and $m_2$ are the magnitudes of the individual
components. This means that the binary system will appear as a single
object with a brighter magnitude than that of the primary
star. Unresolved binaries with MS primary components will thus be
biased towards the brighter envelope of the MS.

We used the \cite{Bres12} isochrones to describe our decontaminated
CMD. Since photometric errors cause the MS to broaden, which hence
renders a robust distinction between single stars and low mass-ratio
binaries difficult, we determined the photometric uncertainties as a
function of magnitude. Following \cite{Hu10}, the relationship is best
described by an exponential function of the form $\sigma(m)=\exp(a m +
b)+c$, where $\sigma$ represents the photometric uncertainty and $m$
is the corresponding magnitude; $a,b$ and $c$ are the fit
parameters.\footnote{These parameters depend only slightly on
  radius, but the variation of $\sigma$ over the range from $V=20$ to
  $V=22$ mag is very small between the cluster centre and the outer
  regions. Since the central cluster annuli contain fewer
  stars than the outer annuli, we checked that using the global
  uncertainty for the generation of a simulated CMD will not introduce
  significant differences in our results.} Figure 4 shows, for NGC
1805, that the functional form of this curve is applicable to our
observations \citep[for NGC 1818, see fig. 3 of][]{Hu11}.

\begin{figure}
 \includegraphics[width=90mm]{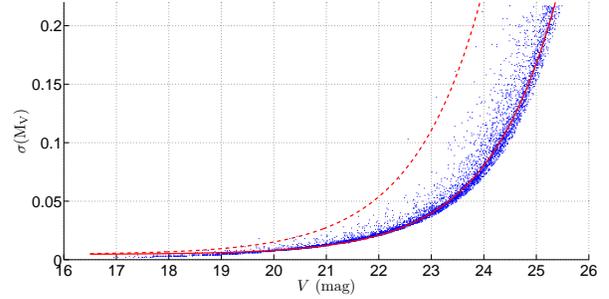}
 \caption{Photometric uncertainties as a function of magnitude for the
   NGC 1805 cluster-centred data set. The red solid line is the
   best-fitting curve to the distribution's ridgeline, while the red
   dashed line represents the 3$\sigma$ range.}
  \label{F4.eps}
\end{figure}

We will now first summarize the fundamental cluster parameters we
adopted for our analysis. Based on isochrone fitting, we obtained a
distance modulus of $(m-M)_0=18.50\;(18.54) \pm 0.02$ mag for NGC 1805
(NGC 1818). Keeping the distance moduli fixed, we determined ages of
$\log(t/{\rm yr})=7.65\pm0.10$ and $7.25\pm0.10$ for NGC 1805 and NGC
1818, respectively. Our derived age for NGC 1805 is identical to that
of \cite{Liu09b}, while the age for NGC 1818 is younger than their
estimate -- $\log(t/{\rm yr})=7.65\pm0.05$ -- but fully consistent
with the age quoted by both \cite{grijs02a}, $\log(t/{\rm
  yr})=7.2\pm0.1$ and \cite{Liu09a}, $\log(t/{\rm
  yr})=7.25\pm0.40$. We note that because of the young age of NGC
1818, age determination based on isochrone fitting results in
significant uncertainties; since \cite{Liu09b} based their age
determination on the faint end of the cluster's MS, it is likely that
their age estimate is an upper limit.

We derive best-fitting extinction values of $E(B-V)=0.09\pm0.01$ mag
for NGC 1805 and $E(B-V)=0.07\pm0.01$ mag for NGC 1818. The former
value is close to the determinations of \cite{Sant13} and
\cite{John01}, $E(B-V)=0.075$ mag. Similarly, our extinction estimate
for NGC 1818 is consistent with that of both \cite{Sant13} --
$E(B-V)=0.07$ mag -- and \cite{John01}, $E(B-V)=0.075$ mag. We adopted
metallicities of $Z=0.008$ and 0.015 for NGC 1805 and NGC 1818,
respectively \citep[cf.][]{Will95,John01}.

For further comparison, \cite{Cast01} and \cite{Liu09b} both adopted
distance moduli of 18.59 and 18.58 mag for NGC 1805 and NGC 1818,
respectively. However, upon close inspection, we note that the
isochrones they adopted describe the blue edge of the cluster MSs,
while our best-fitting isochrones represent the MS ridgelines. This
difference in approach is sufficient to lead to the differences in
basic cluster parameters noted here, i.e., our fits yield
systematically younger ages and lower extinction values than
theirs. We summarize the basic cluster parameters adopted in this
paper in Table 2.

\begin{table}
 \centering
 \begin{minipage}{85mm}
  \caption{Basic parameters of NGC 1805 and NGC 1818.}
  \begin{tabular}{@{}lcc@{}}
  \hline
   Parameter     &  NGC 1805 & NGC 1818 \\
 \hline
 $\log(t/{\rm yr})$ & $7.65\pm0.10$ & $7.25\pm0.10$\\
 $Z$ & 0.008 & 0.015 \\
 $E(B-V)$ (mag) & $0.09\pm0.01$ & $0.07\pm0.01$ \\
 $(m-M)_0$ (mag) & $18.50 \pm 0.02$ & $18.54 \pm 0.02$ \\
 $\alpha_{\rm J2000}$ & $05^{\rm h}02^{\rm m}21.9^{\rm s}$ & $05^{\rm h}04^{\rm m}13.2^{\rm s}$ \\
 $\delta_{\rm J2000}$ & $-66^{\circ}06'43.9''$ & $-66^{\circ}26'03.9''$ \\
  $R_{\rm field}$ (arcsec) & $45.0\pm0.3$ & $72.7\pm0.3$ \\\hline
\end{tabular}
\end{minipage}
\end{table}

\subsection{Binary signatures in the CMD}

We first adopt the MS ridgeline and the region defined by the
3$\sigma$ photometric errors on either side of the best-fitting
single-star isochrone as the area in colour--magnitude space dominated
by single stars. Stars brighter and redder than this limit are treated
as unresolved binary or multiple systems. Binaries with different mass
ratios, $q=M_{\rm s}/M_{\rm p}\leq1$ (where $M_{\rm p}$ and $M_{\rm
  s}$ are the masses of the primary and secondary binary components,
respectively), will cause unresolved binaries to attain magnitudes in
the `binary region', which are brighter than those defined by the
single-star isochrone. We can thus predict where binary systems
characterized by different mass ratios are located in
colour--magnitude space. For the upper limit of the binary region, we
adopt the fiducial binary isochrone for $q = 1$ (representing
equal-mass binaries), i.e., 0.752 mag brighter than the single-star
isochrone, plus a 3$\sigma$ photometric uncertainty, where $\sigma$ is
defined by the spread of the bulk of the single-star ridgeline. Figure
3a of \cite{grijs13} shows an example of the resulting single-star and
binary regions for NGC 1818.\footnote{Note that a precedent exists for
  this approach: similar approaches were also used by \cite{Soll07}
  and \cite{Milo12} for Galactic globular clusters.} We chose to adopt
an upper limit to the binary region to remove some odd objects with
extremely red colours. We speculate that these extremely red objects
may result from possible triple or higher-order systems. Although this
type of stellar system may be rare in regular (open) clusters, triple
and higher-order multiple systems may be observable if they are
associated with very bright stars in massive clusters.

The multiplicity frequency involving triple and higher-order multiple
systems in young massive clusters is unknown.  (In the following,
  when we discuss `multiplicity' we explicitly exclude binary systems
  from our analysis.) In the solar neighbourhood, \cite{Duqu91} found
seven triple systems and two quadruples associated with 164 solar-type
primary stars, implying a $5\pm2$ per cent multiplicity
frequency. \cite{Matt09} simulated star cluster formation and reported
an $18\pm10$ per cent multiplicity frequency for solar-type
systems. In addition, in his table 3 the multiplicity frequency seems
to increase with increasing primary stellar mass. In this paper, we
mainly focus on stars in the range from $V= 20.5\;(20)$ to
$V=22\;(22)$ mag (see below for justification) for NGC 1805 (NGC
1818), roughly corresponding to F-type stars. Also note that
\cite{Matt09} only simulated initial clouds with a mass of 500
M$_{\odot}$, while NGC 1805 and NGC 1818 have masses of
$2.8^{+3.0}_{-0.8}\times10^3$ M$_{\odot}$ and
$2.3^{+1.1}_{-0.3}\times10^3$ M$_{\odot}$, respectively
\citep{grijs02b}.

We hence speculate that these `odd' objects may contain numerous
multiple stellar systems, although some may also be caused by
occasional multiple blends. Since blending of two (artificial) single
stars located along the same line of sight may reach fractions of
order 10 per cent in the clusters' inner regions (see Section 3.4),
the probability of triple or higher-order blending is expected to
reach fractions of order 1 per cent. However, we add the caveat that
possible inadequately decontaminated background stars may also
contribute to their number \citep[for more details regarding the
  decontamination procedure, see][]{Hu10}. It is, nevertheless,
unlikely that such decontamination artefacts could explain the entire
sample of `odd' objects.

Figure 3a of \cite{grijs13} also illustrates why we selected the
magnitude range between $V = 20.5$ and 22 mag for NGC 1805 (and $V =
20$ to 22 mag for NGC 1818) to explore the properties of the clusters'
binaries. In this region of colour--magnitude space, the effects of
binaries are most significant, because the MS slope is shallowest. At
brighter magnitudes, the slopes of both the single-star and equal-mass
binary isochrones are so steep that they lie very close to one
another. In addition, \cite{John01} suggested that the spread in
colour at the brightest extremeties of the MSs of both clusters may
not only be caused by photometric uncertainties, but that blue
straggler stars could also contribute \cite[cf.][]{Li13}. This hence
increases the difficulties associated with investigating binaries at
brighter magnitudes in these distant clusters. For stars fainter than
$V \sim 22$ mag, the increasing photometric uncertainties cause a
significant broadening of the MS, which causes progressively more
binaries to mix with single stars in the CMD, whereas the effects of
incompleteness become appreciable ($\lesssim 50$ per cent
completeness) for $V \gtrsim 24$ mag. In the magnitude ranges probed
for our analysis of the clusters' binary fractions, our sample
completeness is well above 80 per cent (de Grijs et al. 2002a, their
fig. 2; Hu et al. 2010, their fig. 4; see also Section
\ref{AStests.sec}).

\subsection{Cluster centre determination}

Since we focus on exploring the binaries' radial distributions,
accurate cluster centre coordinates are essential. We divided the
stellar spatial distribution into 20 cells in both right ascension
($\alpha_{\rm J2000}$) and declination ($\delta_{\rm J2000}$). We then
proceeded to count the number distributions of stars along both
coordinate directions. A Gaussian-like profile was found to provide
suitable fits to these distributions; the two-dimensional peak of the
adopted Gaussian fits yielded the centre coordinates \citep[see
  also][]{grijs13}. The resulting coordinates are $\alpha_{\rm
  J2000}=05^{\rm h}02^{\rm m}21.9^{\rm s}, \delta_{\rm
  J2000}=-66^{\circ}06'43.9''$ for NGC 1805 and $\alpha_{\rm
  J2000}=05^{\rm h}04^{\rm m}13.2^{\rm s}, \delta_{\rm
  J2000}=-66^{\circ}26'03.9''$ for NGC 1818.

We next calculated the azimuthally averaged number-density profiles of
both clusters (using the full stellar catalogues) and adopted the
distances where the clusters' monotonically decreasing number-density
profiles reach the average field level as the clusters' sizes, i.e.,
$R_{\rm field} = 45.0\pm0.3$ arcsec for NGC 1805 and $R_{\rm field} =
72.7\pm0.3$ arcsec for NGC 1818 \citep{grijs13}. \cite{Mack03}
determined radii for NGC 1805 and NGC 1818 of 69 arcsec and 76 arcsec,
respectively. Their results are based on surface brightness
distributions, so it is reasonable that our results are close but not
exactly the same. Figure 5 shows the number-density profiles of NGC
1805 (top) and NGC 1818 (bottom). In Table 2, we have also included
information about the cluster sizes and their centre coordinates. At
the LMC's distance, for a canonical distance modulus of $(m-M)_0 =
18.50$ mag, 1 arcsec corresponds to 0.26 pc.

\begin{figure}
 \includegraphics[width=95mm]{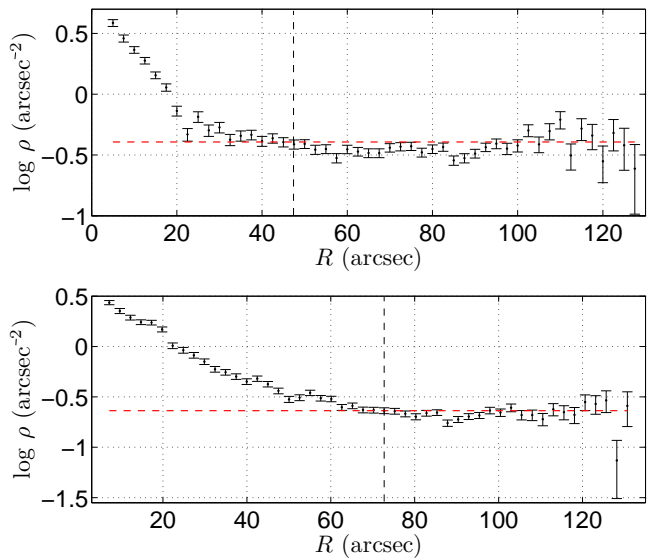}
 \caption{Number-density profiles of (top) NGC 1805 and (bottom) NGC
   1818. The red dashed lines indicate the field density levels, while
   the black dashed lines indicate the distances where the clusters'
   number densities reach these levels, $R_{\rm field}$.}
  \label{F6.eps}
\end{figure}

\subsection{Artificial-star tests}
\label{AStests.sec}

An important effect contributing to broadening of the MS is blending
of otherwise unrelated stars due to crowding and line-of-sight
projection. This causes a similar observational signature as that
owing to the presence of binary systems. To avoid possible
overestimates of the binary fraction caused by stellar blends, we
estimated the blending ratio at different radii using artificial-star
(AS) tests. We adopted the approach of \cite{Hu10}, which was
optimized for application to observations obtained with the {\sl
  HST}/WFPC2 camera, specifically to our observations of NGC
1818. Based on blending experiments of ASs of different luminosities,
they found that two objects which are separated by fewer than 2 pixels
cannot be resolved. We adopted their approach to obtain the blending
fractions at different radii, taking into account that
\begin{enumerate}
  \item stars may blend with each other along the line of sight,
    especially in very crowded environments; and
  \item stellar positions may be coincident with those of cosmic rays
    or extended objects, which may prevent their robust detection.
\end{enumerate}

In general, a large fraction of the brightest objects in our database
are predominantly caused by blending. Since our data set of interest
in this paper is based on a selection that is significantly brighter
than the detection limit, following \cite{Hu10} we added more than
640,000 ASs to the observed, decontaminated stellar catalogues, and
subsequently compared their positions to those of all observed objects
(including stars and extended objects, although we removed possible
cosmic rays from the raw images; this is a negligible effect).  Stars
which are blended with extended sources will be detected as extended
sources and removed from the sample; only stars that actually blend
with other stars will produce a `blending binary', which will cause a
systematic offset of the binary fractions' radial profiles.

We imposed that the ASs obey a luminosity function (LF) corresponding
to a Salpeter-like stellar mass function and that their magnitude and
colour distributions follow the best-fitting isochrones. We then
examined every single AS to check whether it would blend with any
observed object. To do so, we adopted a minimum distance of 2 pixels
between artificial and observed stars as the blending criterion
(cf. Hu et al. 2011, their figs 4--8). If this criterion was met, we
assumed that the fainter of the two objects would be dominated by its
brighter counterpart. We then calculated the ratio of the number of
blended objects (`optical pairs') to the total number of stars in our
sample, $f_{\rm opt}=N_{\rm blend}/N_{\rm tot}$, where $N_{\rm blend}$
and $N_{\rm tot}$ represent the number of relevant objects within a
given magnitude range, and $f_{\rm opt}$ is a function of magnitude.
Figure 6 shows the blending fraction for NGC 1805 as a function of
clustercentric radius and magnitude.

For faint stars, we also need to consider the detection limit. The
observed LF sharply decreases at a given magnitude because of the
onset of sampling incompleteness. In this case, not only stellar
blends affect the detection limit, but a fraction of the fainter stars
may also remain hidden below the detection level. In addition, the
combination of these effects with the complicated structure of the
stellar LF will affect the final shape of the radius-dependent
completeness curve.

To determine our data sets' completeness curves, we generated ASs,
which we first added to the reduced science images and subsequently
recovered using the same photometric approach as employed to obtain
our main stellar databases \citep[cf.][among many others]{Bail92,
  Rube97}. We used {\sc HSTphot} to generate ASs covering the
magnitude range $V \in [19,26]$ mag and $(V-I)$ colours from 0.0 to
1.8 mag, roughly following the distribution of the observational CMD
based on the medium exposure-time images. Note that this fully covers
the magnitude range of interest here, $V \in [20.0,22.0]$ mag, which
we discuss in detail below.

Because of the crowded environments of NGC 1805 and NGC 1818, we would
need to generate more than $10^5$ ASs per image to obtain
statistically robust completeness estimates. However, to avoid blends
between ASs and significant increases of the background's brightness
level, in practice we only added several tens of ASs to our images. In
total, we generated roughly 160,000 ASs for our analysis of both the
NGC 1818 and NGC 1805 {\sl HST} fields, using 70 to 80 stars at a
time. All ASs were randomly distributed across the chips, a process we
repeated 4000 times to reduce statistical fluctuations. Figure 7
displays the 3D `completeness plane' thus obtained. Since both single
stars and binary systems are point sources at the distance of the LMC,
our observations' completeness fractions are simple functions of
radius and magnitude. We thus applied completeness corrections in the
standard manner using the 3D completeness plane of Fig. 7.

In the Appendix, we compare the completeness curves based on both
simulated AS tests \citep{Hu11} and real AS tests (this
paper). \cite{Hu11} reported that their approach can reproduce the
completeness curve for their set of {\sl HST} images. However, we
found that this may only fully apply to sparsely covered images, such
as that represented by the WF3 chip of our NGC 1818 observations
\citep{Hu11}. In significantly more crowded environments, saturated
stars will introduce significant levels of incompleteness which must
be duly taken into account.

\begin{figure}
 \includegraphics[width=90mm]{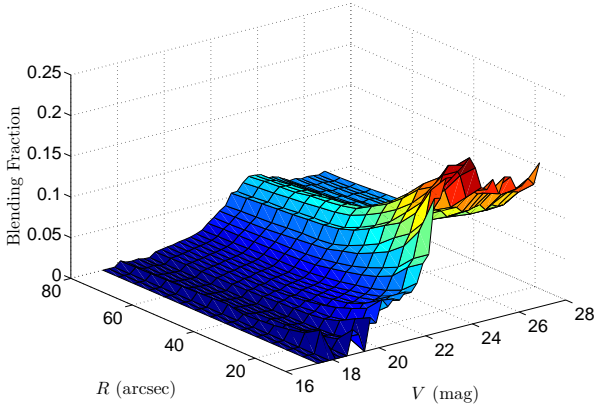}
 \caption{Blending fraction as a function of radius and magnitude for
   NGC 1805.}
  \label{F7.eps}
\end{figure}

\begin{figure}
 \includegraphics[width=90mm]{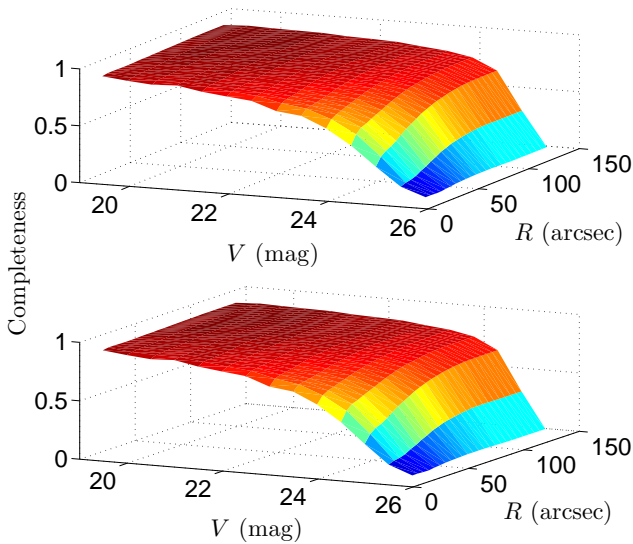}
 \caption{Sample completeness as a function of radius and magnitude
   for (top) NGC 1805 and (bottom) NGC 1818.}
  \label{F8.eps}
\end{figure}

We next calculated the number of stars located in the single-star
region in the CMD divided by that found in the `binary region',
corrected for the corresponding blending fraction of optical pairs,
$f_{\rm opt}$, at different radii for all stars in the magnitude range
of interest for our binarity analysis,
\begin{equation}
f_{\rm bin}=\frac{N_{\rm b}}{(N_{\rm b}+N_{\rm s})}-f_{\rm opt},
\end{equation}
where $N_{\rm b}$ and $N_{\rm s}$ represent the numbers of stars
located in the `binary' and `single-star regions', respectively. We
calculated the resulting `cumulative binary fractions', $f_{\rm bin}$,
for radii from $R \le 10$ to $R \le 45$ arcsec ($R \le 10$ to $R \le
80$ arcsec) in steps of 5 (10) arcsec for NGC 1805 (NGC 1818) and
adopted Poissonian uncertainties. Since the number of stars decreases
as the radius decreases, for $R<10$ arcsec the number of objects is
not statistically significant.

Figure 8 shows the (cumulative) binary fractions' radial profiles for
NGC 1805 and NGC 1818. In the inner regions of both clusters, opposite
trends can be seen. In NGC 1805, the binary fraction first strongly
decreases with increasing radius in the central region, followed by a
slightly and monotonically increasing trend to the cluster's
outskirts, until it reaches the level of the field's binary
fraction. Meanwhile, NGC 1818 displays a monotonically increasing
trend with radius, especially in the innermost regions
\citep[cf.][]{grijs13}. We will discuss the significance of these
results in Section 5.2.

\begin{figure}
 \includegraphics[width=90mm]{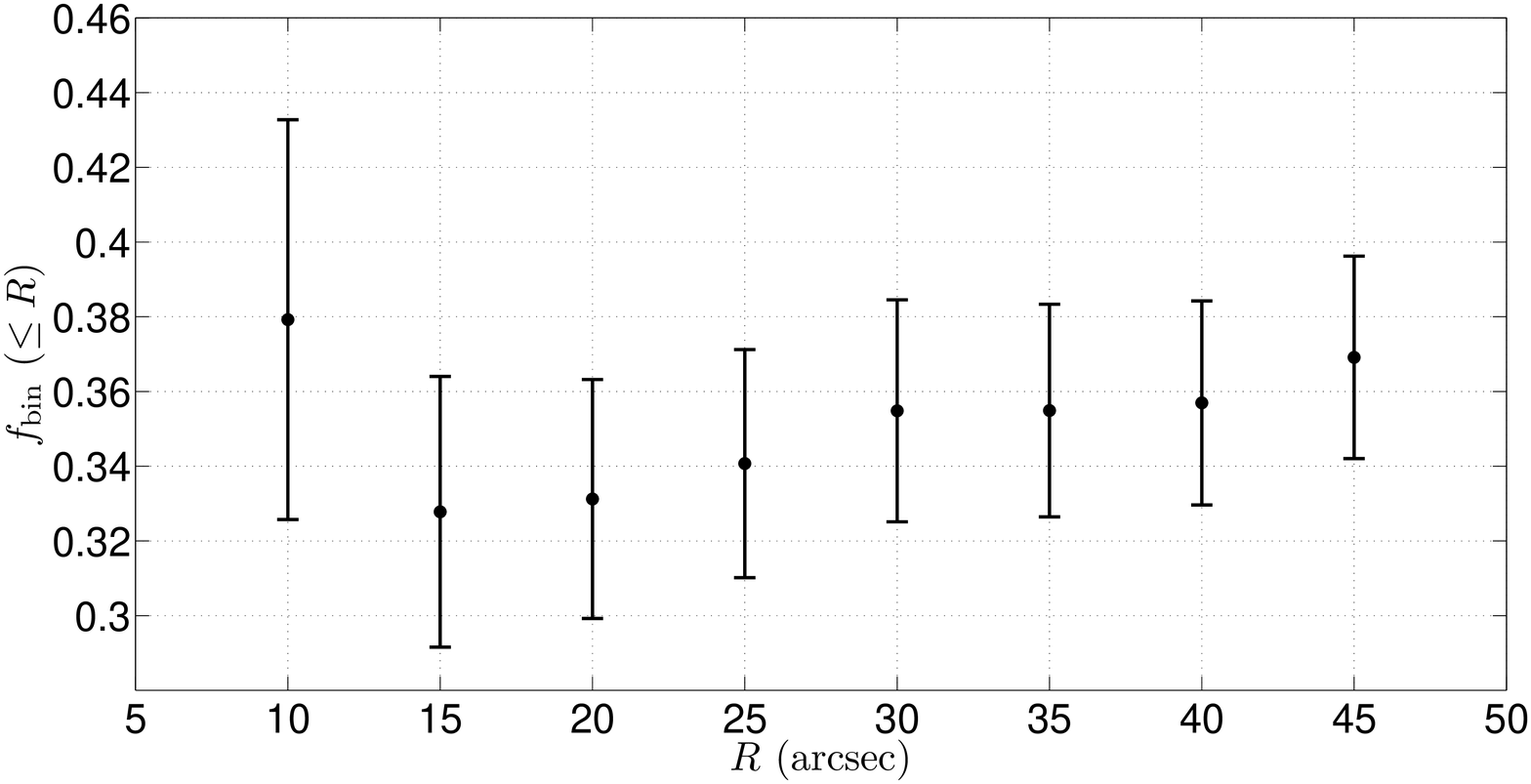}
 \includegraphics[width=90mm]{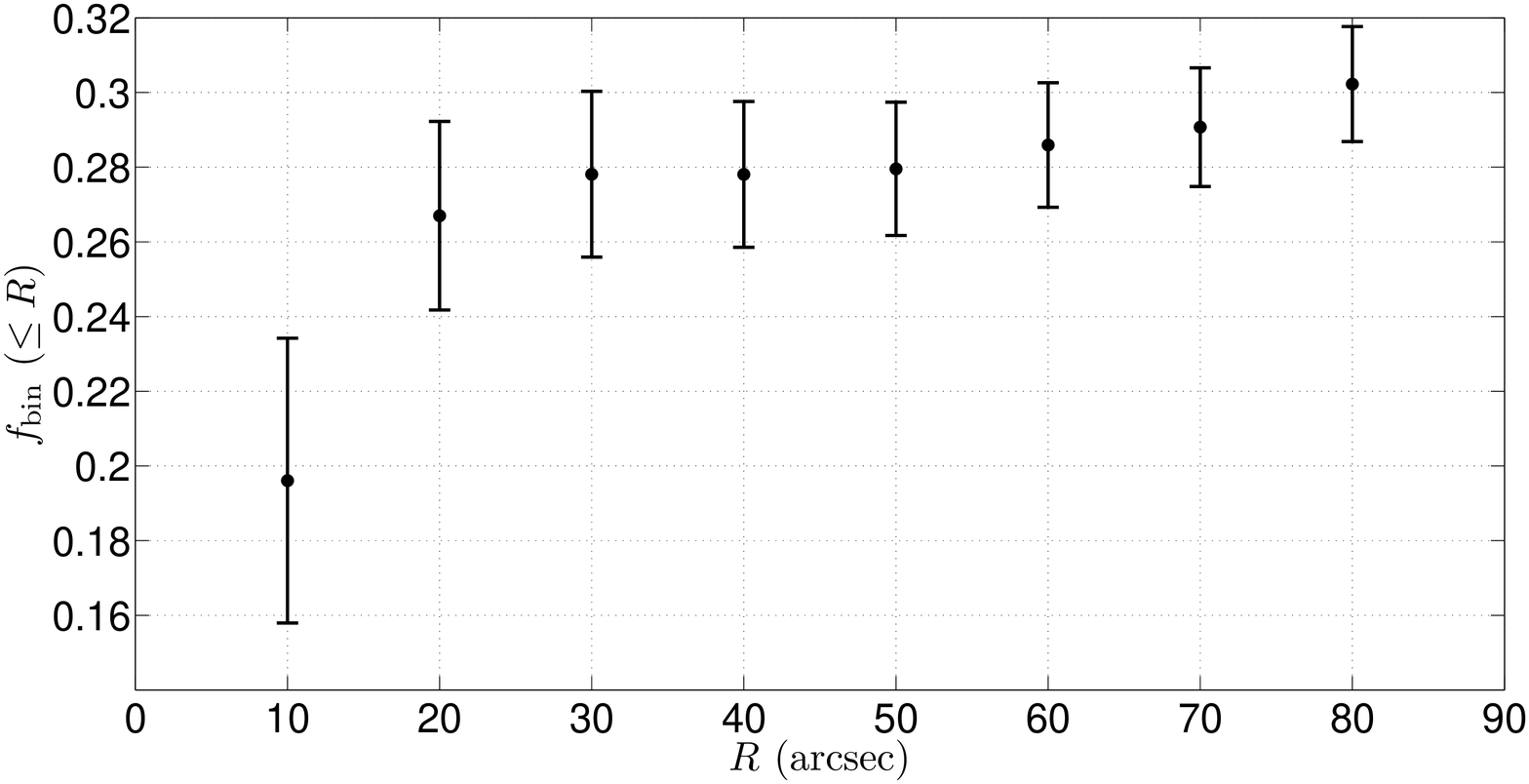}
 \caption{Cumulative binary fraction radial profiles of (top) NGC 1805
   and (bottom) NGC 1818. Error bars represent Poissonian
   uncertainties.}
  \label{F9.eps}
\end{figure}

\section{$\chi^2$ minimization}

We used a more sophisticated and physically better justified approach,
based on Monte Carlo simulations and $\chi^2$ minimization, to test
whether the isochrone-fitting results obtained in the previous section
are robust. Using the best-fitting isochrones, for each WFPC2 chip we
can now simulate a stellar catalogue representing a CMD that is
similar to that of the observed sample. Our AS tests were used to
create millions of ASs. For each chip, we randomly distributed these
across an 800$\times$800-pixel coordinate system. The ASs were drawn
from a Salpeter-like mass function, and the corresponding magnitudes
in both the $V$ and $I$ filters were obtained from interpolation of
the isochrones. Gaussian noise was added to all stars according to the
best-fitting relation based on Fig. 4.

We next transferred the $(x,y)$ pixel coordinates for all stars to the
corresponding $(\alpha_{\rm{J2000}},\delta_{\rm{J2000}})$ values. For
each observed star, we first selected the nearest 20 ASs as possible
couterparts. If the observed object was brighter than $V=20$ mag,
  even if it were an unresolved binary, it would likely mix with the
  genuine single stellar population because of the steepness of the
  CMD: Fig. 9 shows the theoretical ridgelines of single stars and of
  unresolved binary systems characterized by different mass
  ratios. Unresolved binaries fainter than $V=23$ mag can also not be
  distinguished because of the photometric uncertainties affecting our
  data. For such stars, we adopted the AS characterized by the most
similar magnitude and colour as representative counterpart of the
observed star of interest.  For observed stars with magnitudes
  between $V=20$ and $V=23$ mag, we first determined their relative
  positions with respect to the $q = 0.6$ binary population's
  ridgeline. (In our analysis below, we will use a minimum mass-ratio
  cut-off of $q = 0.55$. In view of the uncertainties in $q$, here we
  conservatively choose $q = 0.6$ as the minimum mass ratio.) If an
  observed star is bluer than this locus, we again choose the AS with
  the most similar magnitude and colour as its best-matching
  counterpart. If it is located on the red side of the $q = 0.6$
  boundary but characterized by a colour that is bluer than that of
  the $q = 0.8$ locus, we first calculate the primary star's magnitude
  under the assumption that the observed object is either a $q = 0.6$
  or a $q = 0.8$ binary. We then adopt the average magnitude instead
  of that of the observed object and find the AS with the most similar
  magnitude to that inferred for this primary star. Since all ASs are
  single stars, the ASs that most closely match our observed data
  points hence automatically attain the proper colours for these
  primary stars, {\it modulo} the photometric uncertainties. For
  observed stars that are located between the $q = 0.8$ and $q = 1.0$
  binary loci, we apply a fully equivalent procedure. Stars that are
  located beyond the red side of the $q = 1.0$ binary locus are all
  treated as equal-mass binaries, so that we simply add 0.752 mag to
  both their $V$ and $I$ magnitudes. The resulting stars are hence the
  primary stars of the binary systems that may be contained in our
  photometric database. Figure 10 displays the differences between the
  observed stars and the associated primary stars if we assume the
  observed photometric data points to reflect the colours and
  magnitudes of binary systems. The different colours indicate
  different mass ratios, specifically for $q = 0.6, 0.8$ and 1.0.

\begin{figure}
 \includegraphics[width=95mm]{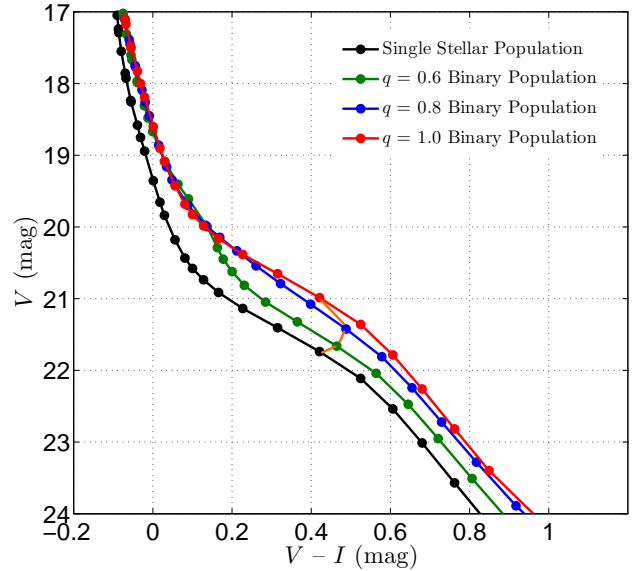}
 \caption{Ridge lines for NGC 1805 of single and binary stellar
   populations characterized by different mass ratios, based on the
   Bressan et al. (2012) isochrones.}
  \label{F12.eps}
\end{figure}

\begin{figure}
 \includegraphics[width=95mm]{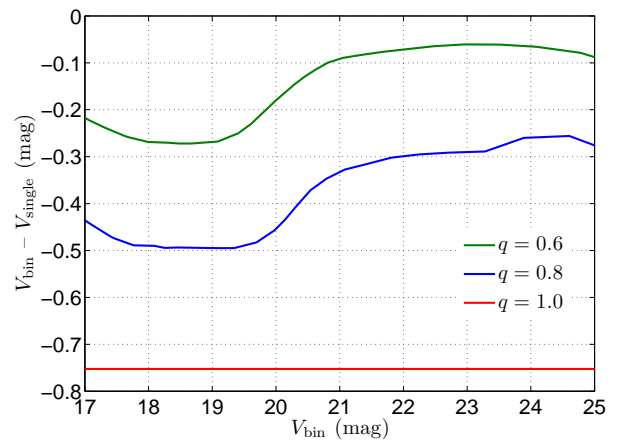}
 \caption{ Magnitude differences between binaries and their
   single-star primaries. Green, blue and red lines represent mass
   ratios of 0.6, 0.8 and 1.0, respectively.}
  \label{F12.eps}
\end{figure}

We thus generated a simulated cluster with an identical number of
stars, a similar spatial distribution to the observed cluster and
  a similar CMD compared with that composed of the observed single
  stars and primary stars of the unresolved binary systems. Again, we
use NGC 1805 as an example to show the comparison between the
simulated and observed clusters. Figure 11 (top) shows the CMD of our
simulated cluster, without inclusion of any binaries or optical pairs
 but instead composed of the binary system's primary stars, and
that of the single-star regime pertaining to NGC 1805. Figure 12 shows
the corresponding spatial distributions. The remaining free parameters
which will cause the distribution of the CMD of the simulated cluster
to differ from the observed CMD are the relevant mass-ratio
distribution of the clusters' binary components and the binary
fraction.

\begin{figure}
 \includegraphics[width=95mm]{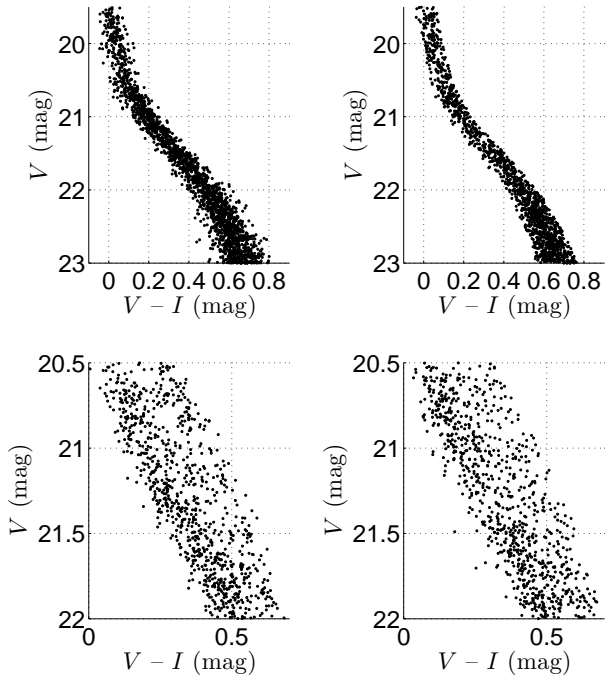}
 \caption{(top left) CMD of the simulated cluster, not including any
   binaries. (top right) Observed CMD of NGC 1805 (`single stars'
   only, i.e., data points found within $\pm 3\sigma$ of the MS
   ridgeline). (bottom left) Simulated cluster CMD, characterized by
   40 per cent binaries and a flat mass-ratio distribution. (bottom
   right) CMD of NGC 1805.}
  \label{F12.eps}
\end{figure}

\begin{figure}
 \includegraphics[width=90mm]{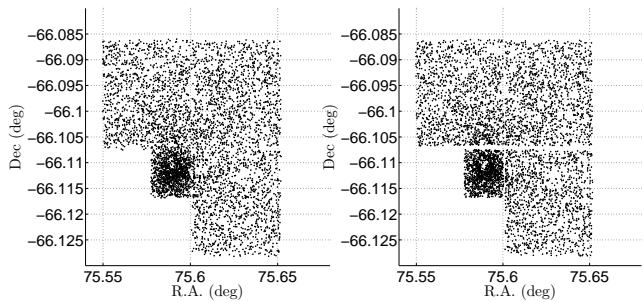}
 \caption{Spatial distributions of (left) our simulated cluster and
   (right) NGC 1805.}
  \label{F11.eps}
\end{figure}

The binary mass-ratio distribution we need to adopt for our
simulations is not set arbitrarily. For binaries in massive clusters,
the most appropriate mass-ratio spectrum is still unknown,
however. \cite{Hu10} chose a flat spectrum (i.e., characterized by a
power-law index, $\alpha = 0$) for NGC 1818, while
\cite{Kouw05,Kouw07} found that a power-law spectrum with an index of
$\alpha = 0.4$ (i.e., low mass-ratio dominated) appears suitable for
binaries in low-density environments. \cite{Regg11} also suggested a
typical range of $\alpha\in[0.0,0.4]$. In fig. 3c of \cite{grijs13},
we showed that if we adopt a proper mass ratio cut-off (here we
initially use $q \ge 0.55$, see below) to mimic the uncertainties
caused by contamination of the binary region by single stars, the
value of the resulting binary fraction does not significantly depend
on $\alpha$ (see below for further improvements). Therefore, we
adopted fixed power-law spectra for the binary mass-ratio distribution
with indices of $\alpha = 0$ and 0.4 for our tests (see below). In
general, the free input parameters are, therefore, the power-law
index, $\alpha$, and the binary fraction, $f_{\rm{bin}}$.

For our analysis of NGC 1805 (NGC 1818), we added artificial binaries
with binary fractions ranging from 5 to 90 per cent, in steps of 5 per
cent, to the selection of observed stars. However, because of
photometric uncertainties, binaries with $q \leq 0.55$ are largely
mixed with the single stars. We thus first adopted a lower mass-ratio
cut-off of $q = 0.55$ for our analysis (but see below); these binaries
will appreciably broaden the MS of the simulated CMD towards brighter
(and redder) photometric measurements.  All stellar photometry
  (including binaries) follows the empirical uncertainties (see
  Fig. 4). Figure 11 (bottom) shows an example of a simulated CMD
(characterized by 40 per cent binaries and a flat-mass ratio spectrum)
and the observed CMD of NGC 1805.  Our simulations are not
  directly corrected for the effects of incompleteness: because we
  determined the incompleteness levels independently and separately,
  we corrected the simulations for these effects upon their
  completion.

We again adopt the same fiducial line as used for our
isochrone-fitting approach to divide the simulated CMD into
single-star and binary regions. For different input binary fractions,
we count how many ASs are located in the single-star region, $N_{\rm
  s}$. These may not all be single stars, because some binaries
characterized by low mass ratios may also be located in this
region. We also count how many ASs are located in the binary region,
$N_{\rm b}$. These stars may not all be binaries either, since this
sample may also contain single stars characterized by a large colour
spread. We take one simulation result of NGC 1805 as an example,
assuming a flat mass-ratio spectrum. The best-fitting model for $R \le
45$ arcsec contains 280 single stars and 125 binaries with $q \ge
0.55$, i.e., $f_{\rm bin}=30.9$ per cent. Further exploration shows
that among these 125 binaries, 12 are located in the single-star
region, while 16 single stars in the simulation have scattered into
the binary region. Our simple isochrone-fitting method would result in
$f_{\rm bin}=31.9$ per cent. The mixture of binaries and single stars
hence results in an overestimation of approximately one per cent based
on the isochrone-fitting approach applied to this cluster. Similarly,
if we adopt a power-law mass-ratio spectrum with an index of 0.4
($\alpha = 0.4$), the mixture will cause an underestimation of 1.4 per
cent. (The number of single stars with a large colour spread remains,
but more low mass-ratio binaries will scatter into the single-star
region.) The larger $\alpha$ is, the higher the number of low
mass-ratio binaries becomes, which hence leads to a larger
underestimation, i.e., $f_{\rm bin(iso)}$ remains constant but $f_{\rm
  bin(\chi^2)}$ increases. The equivalent {\it observed} stellar
numbers derived from our isochrone-fitting approach are denoted
$N'_{\rm s}$ (all treated as single stars) and $N'_{\rm b}$ (all
treated as binaries). We then perform $\chi^2$ minimization:
\begin{equation}
   \chi^2=\left(\frac{N'_{\rm s}-N_{\rm s}}{\sigma_{N_{\rm
         s}}}\right)^2+\left(\frac{N'_{\rm b}-N_{\rm
       b}}{\sigma_{N_{\rm b}}}\right)^2,
 \end{equation}
where $\sigma_{N_{\rm s}}$ and $\sigma_{N_{\rm b}}$ are the
statistical (Poissonian) errors associated with the numbers of ASs, so
that
\begin{equation}
   \chi^2=\frac{(N'_{\rm s}-N_{\rm s})^2}{N_{\rm s}}+\frac{(N'_{\rm
       b}-N_{\rm b})^2}{N_{\rm b}}.
\end{equation}

The $\chi^2$ value indicates the level of similarity between the
simulated and observed CMDs. We vary $f_{\rm{bin}}$ from 5 to 90 per
cent and determine the minimum $\chi^2$ value for each input binary
fraction. Our aim is to determine the global minimum $\chi^2$
value. We therefore use a quadratic function to fit the
$\chi^2(f_{\rm{bin}})$ distribution, which will give us both the
global minimum $\chi^2$ value (and, hence, the best-fitting binary
fraction) and its 1$\sigma$ uncertainty \citep[see
  also][]{grijs13}. The latter corresponds to the difference between
$\chi^2_{\rm{min}}$ and $\chi^2_{\rm{min}}+1$
\citep{Avni76,Wall96}. Figure 13 provides an example of a typical
$\chi^2(f_{\rm{bin}})$ distribution and its corresponding 1$\sigma$
uncertainty. For each radius, we derive $\chi^2_{R}(f_{\rm{bin}})$
using 10 independent generalizations to minimize fluctuations caused
by random scatter. Figure 14 shows the resulting average
$\chi^2_R(f_{\rm{bin}})$ for NGC 1818 for radii from $R\le 10$ to $R
\le 80$ arcsec in cumulative steps of 10 arcsec.  Once we have
  obtained the best-fitting binary fraction from Fig. 14, we correct
  for blending using the results from Fig. 6 (see also the Appendix).

\begin{figure}
 \includegraphics[width=95mm]{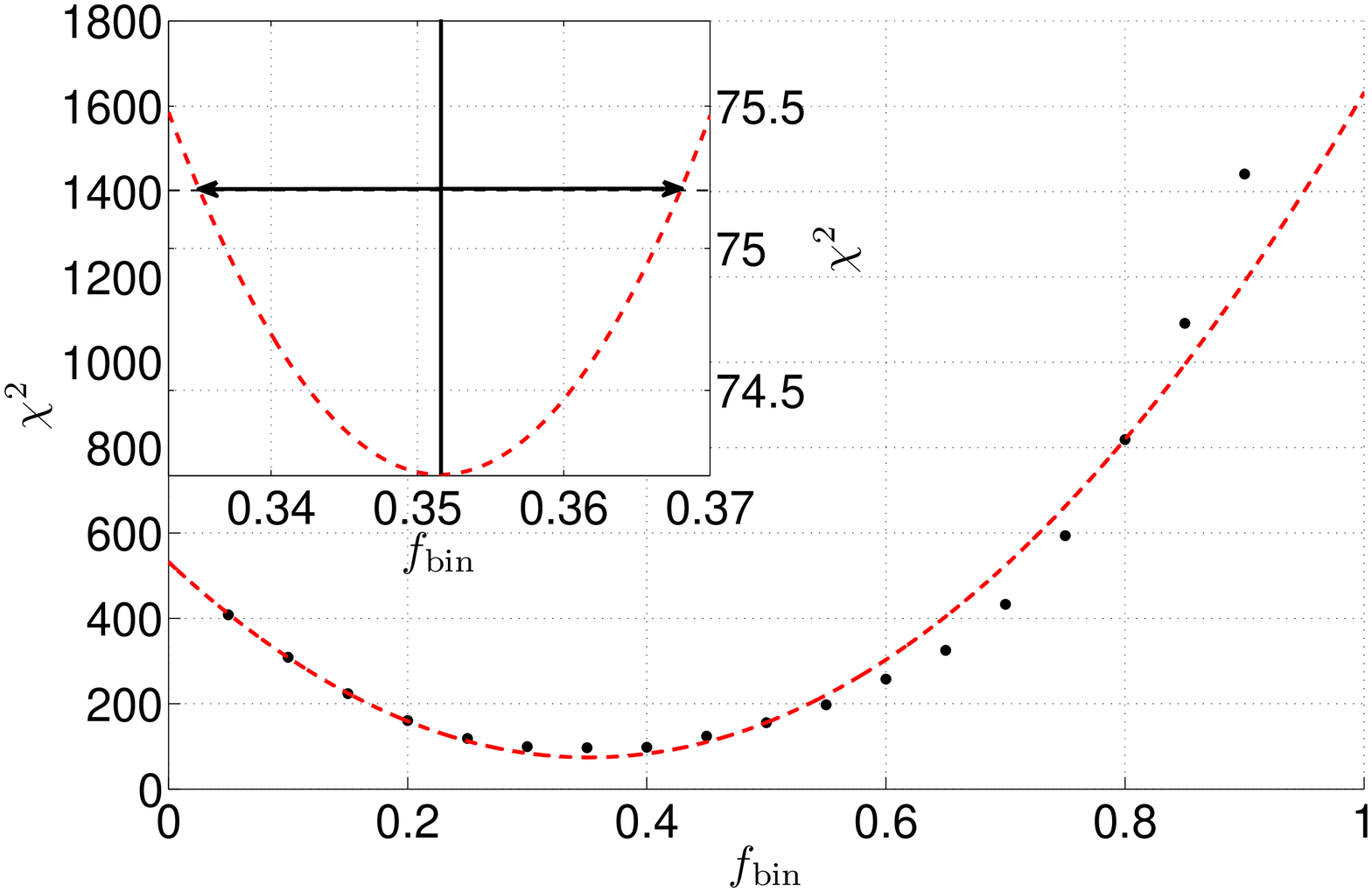}
 \caption{Typical $\chi^2(f_{\rm{bin}})$ distribution. The red dashed
   line is the best-fitting quadratic curve, the double arrow in the
   inset indicates the 1$\sigma$ uncertainty and the vertical black
   line indicates the best-fitting $f_{\rm{bin}}$.}
  \label{F14.eps}
\end{figure}

\begin{figure}
 \includegraphics[width=95mm]{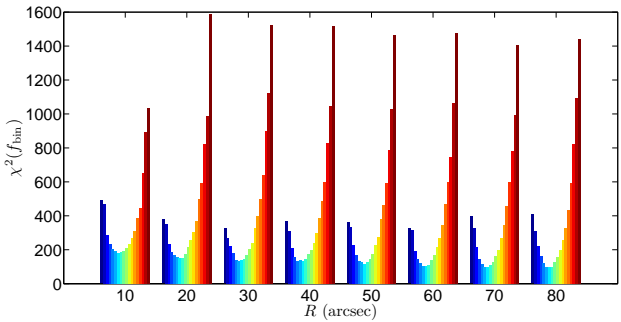}
 \caption{Distribution of $\chi^2_R(f_{\rm{bin}})$ for NGC 1818, from
   $R\le 10$ to $R\le 80$ arcsec. The colour scale represents
   $f_{\rm{bin}}$, from 5 to 90 per cent.}
  \label{F15.eps}
\end{figure}

The $f_{\rm{bin}}(\le R)$ distribution following deblending hence
represents the cumulative binary fraction's radial profile based on
$\chi^2$ minimization. We display the results of the $\chi^2$ tests
for NGC 1805 and NGC 1818 in Fig. 15, where the top and bottom panels
represent NGC 1805 and NGC 1818, respectively (blue and red dashed
lines for mass-ratio distributions assuming, respectively, $\alpha =
0.0$ and 0.4). We also compare the results of our $\chi^2$
minimization with the isochrone-fitting approach (black dashed line)
in the same figure. The comparison shows that both methods mutually
agree very well, which hence implies that both the general behaviour
and the actual levels of the radial profiles of the clusters'
cumulative binary fractions derived here are robust.

Our imposition of a mass-ratio cut-off at $q = 0.55$ is not strictly
necessary to reach convergence. We merely adopted this limit because
we wanted to properly compare the results from both methods. We will
now release this restriction, although in this case we need to make an
informed assumption about the mass-ratio distribution at low mass
ratios. For simplicity and based on the scant observational evidence
at hand (see above), we assume that for all $q$ the mass-ratio
distribution is adequately represented by a flat spectrum, i.e.,
$\alpha = 0.0$. As expected, the result, which is also included in
Fig. 15 (solid blue lines), shows a clearly higher binary fraction
compared to our results for $q \ge 0.55$. However, $\chi^2$
minimization results in the same trend of binary fraction versus
radius as found from our isochrone fits. In fact, this also shows that
the isochrone-fitting approach can only be used to derive what has
been called a `minimum binary fraction' (cf. Sollima et al. 2007;
Milone et al. 2012).

Let us now compare the actual binary fraction determined here with
those resulting from previous work for representative stellar
systems. First, \cite{Soll10} analysed the binary fractions in five
Galactic open clusters using a similar approach as that adopted in
this paper. Adoption of mass-ratio cut-offs of $q \ge 0.48$ to $q \ge
0.55$ leads to core binary fractions between 11.9 and 34.1 per cent in
their five sample clusters. They also estimate a `complete' binary
fraction, i.e., without imposing a mass-ratio cut-off, and find binary
fractions between 35.9 and 70.2 per cent. Clearly, the binary fraction
of NGC 1818 falls within their range of values if we adopt a core
radius of roughly 10 arcsec \citep[see][their table 4]{Mack03}. We
expect that the core binary fraction of NGC 1805 should be
significantly higher than any of the values reported by \cite{Soll10}:
\cite{Mack03} derive a cluster core radius for NGC 1805 of
$5.47\pm0.23$ arcsec, whereas we only explored the binary fraction for
radii in excess of 10 arcsec because of the statistically small number
of stars in the cluster core.

Second, \cite{Milo12} derived the (core) binary fraction of 59
Galactic globular clusters. Their resulting values are significantly
smaller than the values we derive for the young massive clusters NGC
1805 and NGC 1818. They find a maximum core binary fraction of only
roughly 17 per cent, for a mass-ratio cut-off of $q = 0.5$. The fact
that our results yield systematically higher binary fractions than
those reported previously for both globular and open clusters is not a
surprise, not even for the open clusters. In old globular clusters,
the cluster cores will have undergone billions of years of dynamical
processing \citep[this also applies to the intermediate-age open
  clusters of][]{Soll10}, while the binaries in our two young sample
clusters are still in the early stages of their stellar and dynamical
evolution. We will return to the evolution of binary systems in dense
star clusters and the relevant time-scales involved in Section
\ref{binevol.sec}.

\begin{figure}
 \includegraphics[width=90mm]{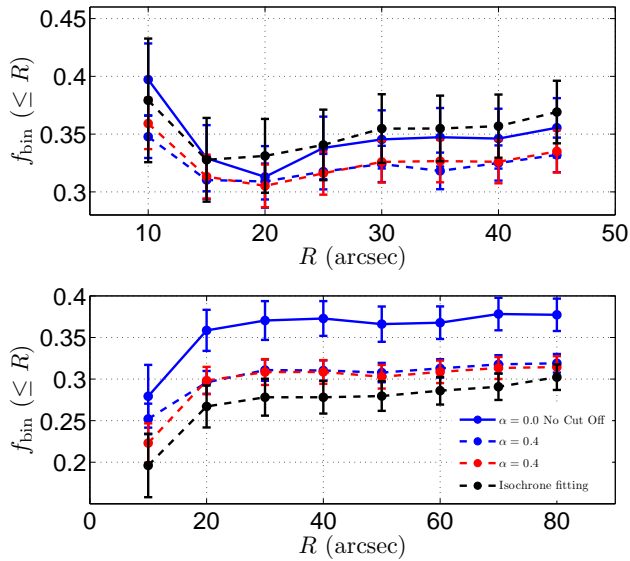}
 \caption{Cumulative binary fraction radial profiles of (top) NGC 1805
   and (bottom) NGC 1818 obtained on the basis of both our $\chi^2$
   tests and isochrone fitting. The blue and red dashed lines
   represent the results from our $\chi^2$ minimization for flat and
   power-law mass-ratio spectra with a $q \ge 0.55$ cut-off
   ($\alpha=0.0$ and $0.4$), respectively; the blue solid lines are
   for $\alpha=0.0$ and no mass-ratio cut-off. The black dashed lines
   indicate the results from our isochrone fits.}
  \label{F16_17.eps}
\end{figure}

We also checked whether the resulting binary fractions depend on
$\alpha$. We found that the choice of $\alpha$ (within observational
limits) does not affect the main trend of the binary fractions' radial
profiles, but it instead determines the relative slope and the value
of the binary fraction at a given radius. This is expected, since
  a steeper mass-ratio distribution (larger $q$) produces more low
  mass-ratio binaries, which will be indistinguishable from single
  stars. In Figs 16 and 17, the 3D surface figures display the
derived binary fraction as a function of both $\alpha$ and radius,
$f_{\rm bin}(\alpha, \le R)$. For both clusters, the main trend
appears stable and robust for $\alpha\leq0.5$. This hence shows that
our choice of $\alpha\in[0.0, 0.4]$ is reasonable.

\begin{figure}
 \includegraphics[width=90mm]{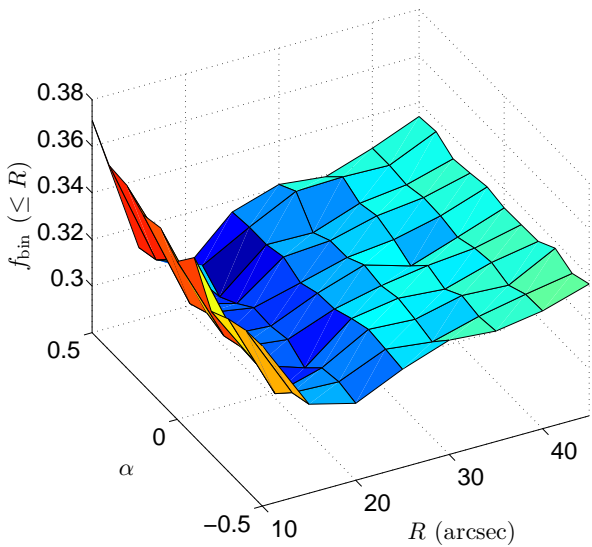}
 \caption{3D surface of $f_{\rm{bin}}(\alpha, R)$ for NGC 1805.}
  \label{F20.eps}
\end{figure}

\begin{figure}
 \includegraphics[width=90mm]{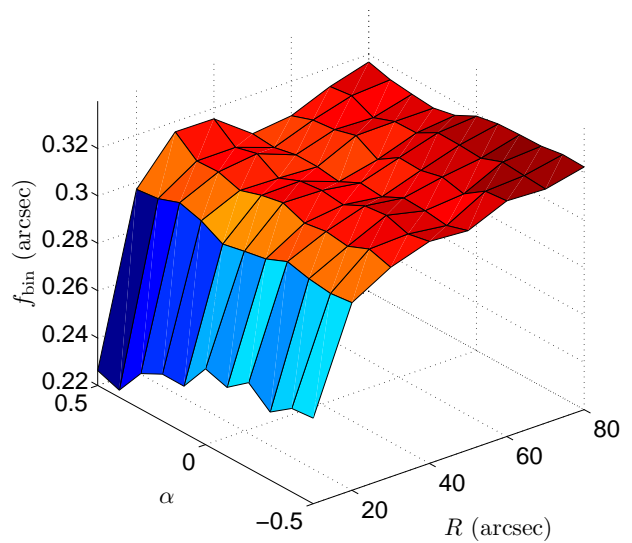}
 \caption{As Fig. 14, but for NGC 1818.}
  \label{F21.eps}
\end{figure}

\section{Discussion}

\subsection{Improvement of previous work}

Following publication of \cite{grijs13}, we significantly improved the
code we developed for $\chi^2$ minimization, which in turn enables us
to obtain more accurate results which are affected by smaller
uncertainties. Our main improvement is related to the way in which we
divide the CMD into cells to calculate $\chi^2(f_{\rm bin})$. In
\cite{grijs13}, we used six cells to cover the colour difference,
$\Delta (V - I)$, adopting best-fitting isochrones as the colour zero
points (hereafter `pseudo-colours'). We also used three cells in the
magnitude direction. However, we found that dividing the CMD in this
manner is not necessary. First, adoption of three cells in magnitude
cannot distinguish any differences between the simulated model and the
observed cluster; the cells in the magnitude direction are only
sensitive to the difference between the adopted and the real LFs.
Since we adopted a fixed LF for our simulated cluster, the cells in
magnitude will contribute nothing but a constant $\chi^2$, and hence
increase the uncertainty.

It also transpired that selection of six cells along the pseudo-colour
axis represented too much subdivision. Since we adopted $q \ge 0.55$
for all binaries, the three bluest cells cannot be used for
distinction of any binaries either. The maximum uncertainty in the
binary fraction reported in \cite{grijs13} reached 9.9 per cent, with
an average uncertainty of 4.4 per cent. For this reason, we needed to
rely on Student's $t$-test scores to estimate at which level our
results were statistically significant. Because of the sizeable
uncertainties, the results from our isochrone-fitting approach could
only be used as additional support.

In this paper, we mimicked the cell-division method applied to our
isochrone fits to proceed with our $\chi^2$-minimization test. In this
case, the results from our isochrone fits can be used both in support
of our conclusion based on the $\chi^2$-minimization approach and as a
comparison method of similar accuracy. Ignoring any uncertainties that
may have been caused by our assumptions of a constant LF and of the
shape of the mass-ratio spectrum at different radii (see Section 5.2
for a discussion), the shape of $f_{\rm bin}(\le R)$ resulting from
$\chi^2$ minimization should be very similar to that based on our
isochrone fitting.

Compared with \cite{grijs13}, the most important improvement we
present here is the significantly improved (i.e., reduced) level of
uncertainty. For instance, for NGC 1818 and assuming $\alpha=0.0\;
(0.4)$, the average uncertainty in the binary fraction is only 1.3
(1.6) per cent. This is significantly smaller than that presented in
our previous work, to the extent that the radial increase of the
binary fraction in NGC 1818 is now clearly statistically robust and we
no longer need to rely on complex statistical tools to assess the
level of significance.

\subsection{Comparison of the approaches}

The isochrone-fitting method is independent of any assumptions as
regards the detailed physical parameters such as the shape of the
stellar mass function and the binary components' mass-ratio
distribution. In addition, NGC 1805 and NGC 1818 are sufficiently
young that we do not need to consider possible contamination from
multiple stellar populations. Introduction of multiple stellar
populations becomes important only for intermediate-age massive star
clusters, many of which exhibit a significant double or extended MS
turnoff \citep{Mack07,Mack08,Gira09,Milo09}. 

Some authors have suggested that double or extended MS turnoffs in
intermediate-age massive clusters may be owing to rapid stellar
rotation \citep[e.g.,][]{Bast09,Li12}, because the resulting
centrifugal support will decrease the effective temperatures and
luminosities of the stars. This fast-rotation effect may not only
affect intermediate-age but also young massive clusters. To estimate
the possible effect of fast rotation in the context of our target
clusters, we follow \cite{Bast09} and adopt a rotation rate of $\omega
= 0.55$ (expressed as fraction of the critical break-up rotation rate)
and a standard deviation of 0.15. We also assume that $\omega$
increases linearly with mass from 1.2 M$_{\odot}$ to 1.5 M$_{\odot}$
\citep[cf.][]{Li12}. In intermediate-age massive clusters, the
reddening caused by rapid rotation is quite significant for stellar
masses close to 1.5 M$_{\odot}$ (corresponding to apparent magnitudes
at the distance of the LMC near $V = 20$ mag). On the other hand, for
our target clusters this effect is trivial, because at intermediate
ages such stars are found near the MS turnoff while in young massive
clusters they are still located on or very near the zero-age MS and
attain very blue (hot) colours. The temperature decrease caused by
rapid rotation is insignificant compared with the temperature of
non-rotating stars.

Taking NGC 1805 as an example, Fig. 18 shows the best-fitting
isochrones for no and fast rotation. The two ridgelines are quite
close, characterized by an offset of only approximately 0.02 mag in
colour. We therefore added 0.02 mag to the photometric uncertainties
of our sample stars in the relevant magnitude range. The minimum age
for which a distinct effect of rotation on the MS locus of massive
clusters would show up is still unclear. \cite{Yang13} suggest that,
at least for massive clusters younger than 0.6 Gyr, the effects of
rotation are insignificant.

\begin{figure}
 \includegraphics[width=90mm]{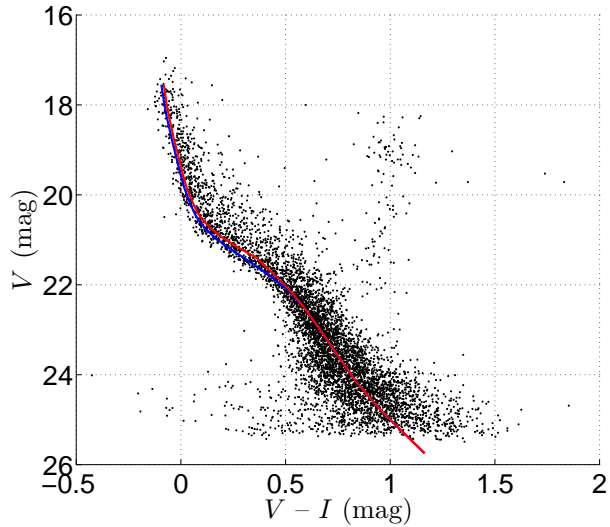}
 \caption{Effect of rapid stellar rotation in NGC 1805. The blue line
   represents the best-fitting isochrone without considering rapid
   rotation; the red line is for rapidly rotating stars ($\omega =
   0.55$ of the critical break-up rotation rate). Only a small offset
   is seen for $V\in[20, 22]$ mag. }
  \label{F22.eps}
\end{figure}

More importantly, however, isochrone fitting is limited by the
necessity to impose a minimum binary fraction in our analysis. Because
the intrinsic colour spread of single stars caused by photometric
uncertainties broadens the MS, only stars that are found significantly
beyond the MS in colour--magnitude space can be treated robustly as
binaries. In addition, Poissonian uncertainties will further reduce
the number of useful objects in the innermost regions of dense
clusters.

Alternatively, our newly developed $\chi^2$ minimization approach can,
in principle, be used to determine binary fractions more accurately
and for a larger range of mass ratios. In the context of isochrone
fitting, the uncertainties are driven by the total number of stars, so
that the absolute errors increase with decreasing radii. For instance,
for NGC 1805, isochrone fitting (assuming a flat mass-ratio
distribution) yields uncertainties that increase from 2.5 to 5.4 per
cent as radii decrease from 45 arcsec to 10 arcsec. However, the
uncertainties associated with our $\chi^2$ minimization increase only
from 1.5 to 1.8 per cent for the same radial range (a power-law
mass-ratio distribution also returns uncertainties of better than 2
per cent), because the absolute uncertainties only depend on the slope
of $\chi^2(f_{\rm{bin}})$ as a function of radius (cf. Fig. 14).

However, the main disadvantage of our $\chi^2$-minimization approach
is that it takes a relatively large amount of computer time compared
to isochrone fitting. For example, for each of our tests we ran
simulations using roughly 2 million ASs. A single simulation takes
approximately 3 min in a normal core i7 CPU environment. For each
cluster, we ran the simulations 10 times in eight radial bins to
determine the average $\chi^2(f_{\rm bin})$ distribution for a given
radial range. This hence consumes around 4 hr. Our tests in which we
varied the mass-ratio distribution from $\alpha = -0.9$ to $+0.9$ in
steps of $\Delta \alpha = 0.1$ showed that the total computer time
required for one cluster was $4\times19=76$ hr. Using standard
multiple processors, these tests would still take of order 2 days to
complete.

In addition, for the $\chi^2$ tests, we need to assume a mass function
and a mass-ratio spectrum for the binary population. However, the mass
function is likely to vary as a function of radius because of the
effects of mass segregation. At the same time, binaries characterized
by low mass ratios in the clusters' inner regions will be
preferentially disrupted, so that the mass-ratio distribution will
also vary as a function of radius. For instance, from Fig. 14 it is
clear that even though the $\chi^2(f_{\rm{bin}})$ distributions at
different radii are all quite smooth, the distribution for $R \le 10$
arcsec is systematically and significantly offset to higher mass
ratios compared with the other radii. This means that the assumption
of a uniform mass function and a constant mass-ratio spectrum for the
entire cluster is inappropriate. However, since the shape of the
binary fraction's radial profile does not vary very significantly, we
nevertheless adopted a simple, single mass function and mass-ratio
distribution to limit the total amount of computing time needed for
our simulations.

\subsection{Physical implications}

\subsubsection{Binary segregation}
\label{binevol.sec}

The cumulative binary fraction's radial profile in NGC 1805 displays a
significant decreasing trend for the innermost two radial bins ($R
\le15$ arcsec), followed by a slight increase to the field's binary
fraction. Since binaries are, on average, more massive than single
stars of similar spectral types, it is reasonable to conclude that
mass segregation is likely responsible for the behaviour of the radial
binary fraction in NGC 1805.

\cite{grijs02b} reported a clear detection of the effects of mass
segregation in NGC 1805, although these authors assumed that all stars
in their sample were single stars. They pointed out that it is not
straightforward to correct the observed LF for the presence of
binaries, in particular since the binary fraction as a function of
brightness is difficult to determine. Note that, despite significant
progress and improvements, here we still can only determine the binary
fraction for cluster MS stars spanning a narrow mass range (roughly
for $m_{*}\in[1.3 \; {\rm M}_{\odot}, 1.9 \; {\rm M}_{\odot}]$,
corresponding to F-type primary stars).

To understand the radial dependence of the binary fraction of F-type
stars in NGC 1805, we need to compare the time-scale governing early
dynamical mass segregation with the cluster's age. \cite{grijs02b}
calculated that the core of NGC 1805 is 3--4 crossing times old,
although they adopted a chronological age of $\log( t/{\rm yr}) =
7.0^{+0.3}_{-0.1}$. Since the age we adopted is older, this implies
that the cluster age in units of the crossing time reported in
\cite{grijs02b} is a lower limit. In fact, adopting an age of $\log(
t/{\rm yr}) = 7.65\pm 0.10$, we estimate that the core of NGC 1805 may
be of order 8--10 crossing times old. In addition,
\cite{Alli09,Alli10} pointed out that if -- as suggested by
observations -- clusters form clumpy rather than as spherically
homogeneous Plummer spheres \citep[see also][]{Moec09}, they will
undergo early dynamical mass segregation more quickly, on time-scales
of 1--2 Myr for the most massive stars in the cores of their open
cluster-like model systems \citep[for an alternative scenario leading
  to rapid mass segregation, see][]{McMillan12}. In view of these
time-scale arguments and compared with the cluster's chronological
age, its binary population should have been modified by dynamical
interactions.

Binary systems can be formed through both three-body encounters (`3b')
and tidal capture \citep[`tc';][]{Spit87}, i.e.,
\begin{equation}
  \left(\frac{{\rm d}n_{\rm b}}{{\rm d}t}\right)_{\rm
    tot}=\left(\frac{{\rm d}n_{\rm b}}{{\rm d}t}\right)_{\rm
    3b}+\left(\frac{{\rm d}n_{\rm b}}{{\rm d}t}\right)_{\rm tc}.
\end{equation}
\cite{Spit87} provides a useful numerical approximation to the
formation rate of binaries through three-body encounters,
\begin{equation}
\begin{aligned}
  \left(\frac{{\rm d}n_{\rm b}}{{\rm d}t}\right)_{\rm 3b}=&
  1.97\times10^{-13}\left(\frac{n}{10^4\;{\rm
      pc}^{-3}}\right)^3\left(\frac{m_*}{{\rm
      M}_{\odot}}\right)^5\\ &\times\left(\frac{10\mbox{ km
      s}^{-1}}{\sigma_{ m_*}}\right)^9{\rm pc}^{-3}{\rm yr}^{-1},
  \end{aligned}
\end{equation}
where $n$ and $\sigma_{m_*}$ are the stellar density and 3D velocity
dispersion of single stars, respectively. For the rate of binary
formation through tidal capture, we have
\begin{equation}
\begin{aligned}
  \left(\frac{{\rm d}n_{\rm b}}{{\rm d}t}\right)_{\rm tc}=&
  10^{-8}\kappa\left(\frac{n}{10^4\;{\rm
      pc}^{-3}}\right)^2\left(\frac{m_*}{{\rm
      M}_{\odot}}\right)^{1+\mu/2}\\ 
&\times\left(\frac{R_{\rm s}}{{\rm
      R}_{\odot}}\right)^{1-\mu/2}\left(\frac{10\mbox{ km
      s}^{-1}}{\sigma_{m_*}}\right)^{1+\mu}{\rm pc}^{-3}{\rm yr}^{-1}.
 \end{aligned}
 \end{equation} 
Here, $(\kappa,\mu)=(1.52,0.18)$ and (2.1,0.12) for the polytropes $n
= 3$ (resulting from the Eddington standard model of stellar structure
for MS stars) and 1.5 (relevant to degenerate stellar cores, white
dwarfs and less massive bodies), respectively, and $R_{\rm s}$ is the
stellar radius.

Let us now estimate the number of binaries that may have formed
through tidal capture during the lifetime of NGC 1805. \cite{Mack03}
determined a (central) density of $\log\rho_0 [{\rm M}_{\odot}\;{\rm
    pc}^{-3}] = 1.75\pm0.06$ and a total mass of $\log(M_{\rm cl}[{\rm
    M}_{\odot}]) = 3.45^{+0.10}_{-0.11}$ for this cluster.  Within a
radius of $45''$, which roughly corresponds to the size of NGC 1805
(see Table 2), we detected 3500 stars.  Corrected for
  incompleteness,\footnote{The completeness level sharply
    decreases near $V = 24.0$ mag. For fainter stars, we assume a
    Kroupa (2001)-like LF shape down to the hydrogen-burning limit at
    0.1 M$_\odot$. This shape is, to first order, appropriate for both
    NGC 1805 and NGC 1818 (cf. Liu et al. 2009a,b). we estimate that
  NGC 1805 contains a total of approximately 10,000 stars}. We now
simply assume that this sample represents the total number of cluster
members, so that the average stellar mass in the cluster is of order
0.28 M$_{\odot}$.\footnote{ It is encouraging -- and instills
  confidence in our results -- that the cluster masses implied by
  simply adding up the individual stellar masses obtained from
  integrating both clusters' stellar mass functions over the full mass
  range, including the uncertain extrapolation to lower stellar
  masses, are similar to the cluster masses reported in the
  literature, at least to first order. Here, we derive cluster masses
  for NGC 1805 and NGC 1818 of $\log(M_{\rm cl}/{\rm M}_\odot) \sim
  3.8$ and 4.0, respectively, which must be compared with $\log(
  M_{\rm cl}/{\rm M}_\odot) = 3.45^{+0.10}_{-0.11}$ and $4.01 \pm
  0.10$, respectively, from \cite{Mack03}.} This corresponds to a
central number density of $\rho = 70$ stars pc$^{-3}$, assuming that
the average mass of stars for $V \in [20.5, 22.0]$ mag is
1.6~M$_{\odot}$. MS stars in this mass range have radii of roughly
1.3~R$_{\odot}$; the cluster's stellar velocity dispersion is still
unknown. \cite{grijs02b} estimated that the velocity dispersion of NGC
1805 should be more than 10 times smaller than that of NGC 1818, while
\cite{Elso87b} estimated a velocity dispersion for NGC 1818 of $\geq
6.8$ km s$^{-1}$. We hence simply assume a velocity dispersion for NGC
1805 of $\sim 0.7$ km s$^{-1}$.  Also, we adopt 5 arcsec as projected
core size of NGC 1805, which roughly corresponds to 1.3 pc at the
distance of the LMC. These considerations imply that, by the cluster's
current age, only 5.6 and 0.2 binaries are expected to have formed in
the cluster core through three-body encounters and tidal capture,
respectively.

NGC 1805 exhibits a significantly decreasing binary fraction out to
$R=15$ arcsec. Since \cite{Mack03} determined a cluster core radius of
$5.47\pm0.23$ arcsec, this decreasing trend covers up to 3 core
radii. We do not explore the binary fraction in the innermost 5 arcsec
because of the large uncertainties associated with the small number of
stars in that region. Nevertheless, it is still reasonable to conclude
that the higher binary fraction in the cluster's inner region is most
likely due to the effects of early dynamical mass segregation
\citep[cf.][]{Alli09,Alli10}, a process akin to violent
relaxation. After all, binary systems are, on average, more massive
than the single stars they are composed of, so that they are subject
to a more significant degree of dynamical friction and, hence, mass
segregation.

In addition, as the more massive stars sink towards the cluster
centre, their dynamical evolution speeds up \citep{Gurk04}. This will
be accelerated if -- as usual in realistic star clusters -- there is
no full energy equipartition \citep{Inag85}, thus producing
high-density cores very rapidly, where stellar encounters occur very
frequently and binary formation is thought to be very effective
\citep{Inag85,Elso87b}. In fact, the presence of binary stars may
accelerate this early dynamical mass segregation significantly, since
two-body encounters are very efficient
\citep{Neme87,Marc96,Bonn98,Elso98,Park09}. This process will act on
similar (or slightly shorter) time-scales as conventional dynamical
mass segregation, which occurs through standard two-body relaxation.

\subsubsection{Binary dynamical disruption}

Since NGC 1818 has a similar age as NGC 1805, the opposite behaviour
of the radial binary fraction seen for NGC 1818 indicates that
additional dynamical effects may also play an important role. In
\cite{grijs13}, we concluded that the apparent deficit of binaries in
the NGC 1818 core may be the observational signature of the
preferential disruption of soft binary systems. In this paper, our
improved $\chi^2$-minimization procedure has yielded an even more
robust increasing trend for NGC 1818's radial binary fraction than
that reported by \cite{grijs13}. This thus underscores our earlier
conclusion.

In massive star clusters, hard binaries get harder, on average, and
soft binaries become softer \citep{Hegg75,BT87}. In addition,
\cite{BT87} point out that the time-scale governing soft binary
systems should always be significantly shorter than the local
relaxation time-scale. This implies that if there are sufficient
numbers of soft binary systems and the cluster has undergone evolution
for longer than its half-mass relaxation time, binary disruption
should have proceeded efficiently.

\cite{grijs02b} calculated an age of NGC 1818's core of $\geq$ 5--30
crossing times, which compares favourably with our age estimate of the
NGC 1805 core, also in units of crossing times. Both clusters have
similar chronological ages, yet their binary fractions' radial
profiles are markedly different. What causes this apparent diversity?
Disruption of binary systems is driven by kinetic-energy transfer from
a cluster's bulk stars to the binary members of interest.  Once the
velocities of the binary components are much lower than those of the
bulk stellar population, such a binary system can be treated as
soft. \cite{BT87} derived the `watershed' energy of soft binaries as
$-m\sigma^2$, where $\sigma$ is the local velocity dispersion of the
environment in which the binaries reside. Since the mass ranges of the
binaries we analysed in both NGC 1805 and NGC 1818 are similar, the
main factor which determines if a binary system is a soft binary
system is its velocity dispersion. Since \cite{grijs02b} suggested
that the velocity dispersion of the NGC 1818 core may be roughly 10
times greater than that of the core of NGC 1805, many more binaries in
NGC 1818 are thought to be soft binaries. They are more easily
disrupted in the dense core of NGC 1818, while such binary systems are
likely to survive more easily in NGC 1805. In addition, NGC 1818 also
has a relatively denser core compared with NGC 1805. Sollima (2008;
his fig. 6) derived a relationship between the survival frequency of
binary systems and the density and velocity dispersion of their host
clusters, which shows that binaries survive much more easily in
environments characterized by relatively low densities and velocity
dispersions. We speculate that these differences in physical
conditions are at the basis of the significantly different radial
binary fraction profiles observed for both clusters.

Let us now adopt velocity dispersions for NGC 1805 and NGC 1818 of 7
and 0.7 km s$^{-1}$, respectively, and total masses of $\log( M_{\rm
  cl}[{\rm M}_{\odot}]) = 3.45^{+0.10}_{-0.11}$ and $4.01\pm{0.10}$.
 Corrected for incompleteness, we concluded that of order 10,000
  stars are likely associated with NGC 1805 within a radius of 45
  arcsec; for NGC 1818 we estimate a total stellar membership count of
  14,500 stars within 73.7 arcsec. We hence adopt the simple assumption
  that NGC 1805 (NGC 1818) is composed of 10,000 (14,500) stars, which
  allows us to derive an average stellar mass of $\bar{m} =
  0.28^{+0.07}_{-0.06} (0.71^{+0.18}_{-0.15})$ M$_{\odot}$. These
values allow us to estimate the typical semi-major axis length of the
binary systems in both clusters,
\begin{equation}
   a=\frac{G\bar{m}}{2\sigma^2}
\end{equation}
where $G$ is the gravitational constant and $\sigma$ is the cluster's
velocity dispersion. For binary systems to survive for a significant
period of time in the cluster centre, the typical minimum semi-major
axis required for NGC 1805 is of order  250 au, while that for
NGC 1818 is only  $\sim 6$ au. \cite{Duqu91} systematically
explored the properties of the binarity and higher-order multiplicity
of stars in the solar neighbourhood. they found for nearby field
binary systems that the distribution of their typical semi-major axes
peaks at 30 au. If we adopt the simplistic assumption that the same
solar-neighbourhood conditions also apply to the LMC field, this may
indicate a medium value for the semi-major axes of surviving binaries.
  
These arguments thus support the notion that relatively wide binaries
are more easily disrupted in the environment of the NGC 1818
core. This also corroborates the idea that dynamical evolution is
likely to have acted on the stars in the core of NGC 1818, but also in
NGC 1805 (although to a lesser degree). In turn, this is underscored
by the results of \cite{grijs02a,grijs02b}, who found significant mass
(luminosity) segregation of the single stars in both clusters out to
several core radii. Note that these authors did not correct their
results for binarity, although the bulk ($\gtrsim 60$ per cent) of
their sample stars would have been single stars. Combined with recent
insights into rapid dynamical evolution in initially cool cluster
cores \citep[e.g.][]{Alli09,Alli10}, this suggests that at least some
of the observed degree of mass segregation may be dynamical in origin.

\section{Conclusions}

In this paper, we have used isochrone fitting and $\chi^2$
minimization to investigate the binary fractions as a function of
radius in the young massive LMC star clusters NGC 1805 and NGC
1818. Both methods agree very well with one another as regards the
radial trends of the clusters' binary fractions. This implies that
both approaches yield robust results.

Our scientific results exhibit opposite trends regarding the behaviour
of the binary fractions as a function of radius in the inner regions
of both clusters. For NGC 1805, we detected a significant decrease in
the binary fraction from the inner core to the cluster's
periphery. Meanwhile, for NGC 1818 we found a monotonic increase of
the binary fraction with radius. We conclude that while early
dynamical mass segregation and the disruption of soft binary systems
should be at work in both clusters, time-scale arguments imply that
early dynamical mass segregation should be very efficient and, hence,
likely dominates the dynamical processes in the core of NGC
1805. Meanwhile, in NGC 1818 the behaviour in the core is probably
dominated by disruption of soft binary systems. We speculate that this
may be owing to the higher velocity dispersion in the NGC 1818 core,
which creates an environment in which the efficiency of binary
disruption is high compared with that of the NGC 1805 core.

\section*{Acknowledgments}

We are grateful for support from the National Natural Science
Foundation of China through grants 11073001 and 10973015.


\section*{Appendix}
\label{sec:app}

In this Appendix, we further compare the quality and accuracy of the
completeness curves for both NGC 1805 and NGC 1818 obtained using both
simulated AS tests based on catalogue data \citep{Hu10,Hu11} and
`real' (actual) AS tests as employed in this paper. We added `real'
ASs to the reduced images using the {\sc HSTphot} package, which can
generate ASs that resemble real objects as observed with the {\sl
  HST}/WFPC2 camera.

\cite{Hu11} claimed that on the basis of their catalogue-based method,
they can very well reproduce the NGC 1818 completeness curve, as if
they had employed actual AS tests. However, they only show the
completeness curve for the WFPC2/WF3 chip. On close inspection, we
found that their results may only be correct for the relatively
low-density environments for which they carried out their
comparison. We repeated their catalogue-based and our actual AS tests
using our NGC 1805 data, and indeed found excellent agreement between
both completeness curves in the low-density regions on the WFPC2 (WF3)
chips.

However, if we consider more crowded regions, the incompleteness
levels resulting from the actual AS tests are significantly lower than
those from the catalogue-based method of \cite{Hu11}. Figures 19 and
20 show the completeness curves of the four WFPC2 chips for the NGC
1805 and NGC 1818 data, respectively. This result implies that the
method proposed by \cite{Hu11} may only be fully applicable to
low-density regions and a one-to-one correspondence breaks down above
a certain density threshold. (The actual density threshold for which
this statement holds requires in-depth exploration of the data at
magnitudes well outside the range considered here for our binarity
analysis. This is beyond the scope of the present paper, but will be
explored in a future, technical contribution.)

\begin{figure}
 \includegraphics[width=90mm]{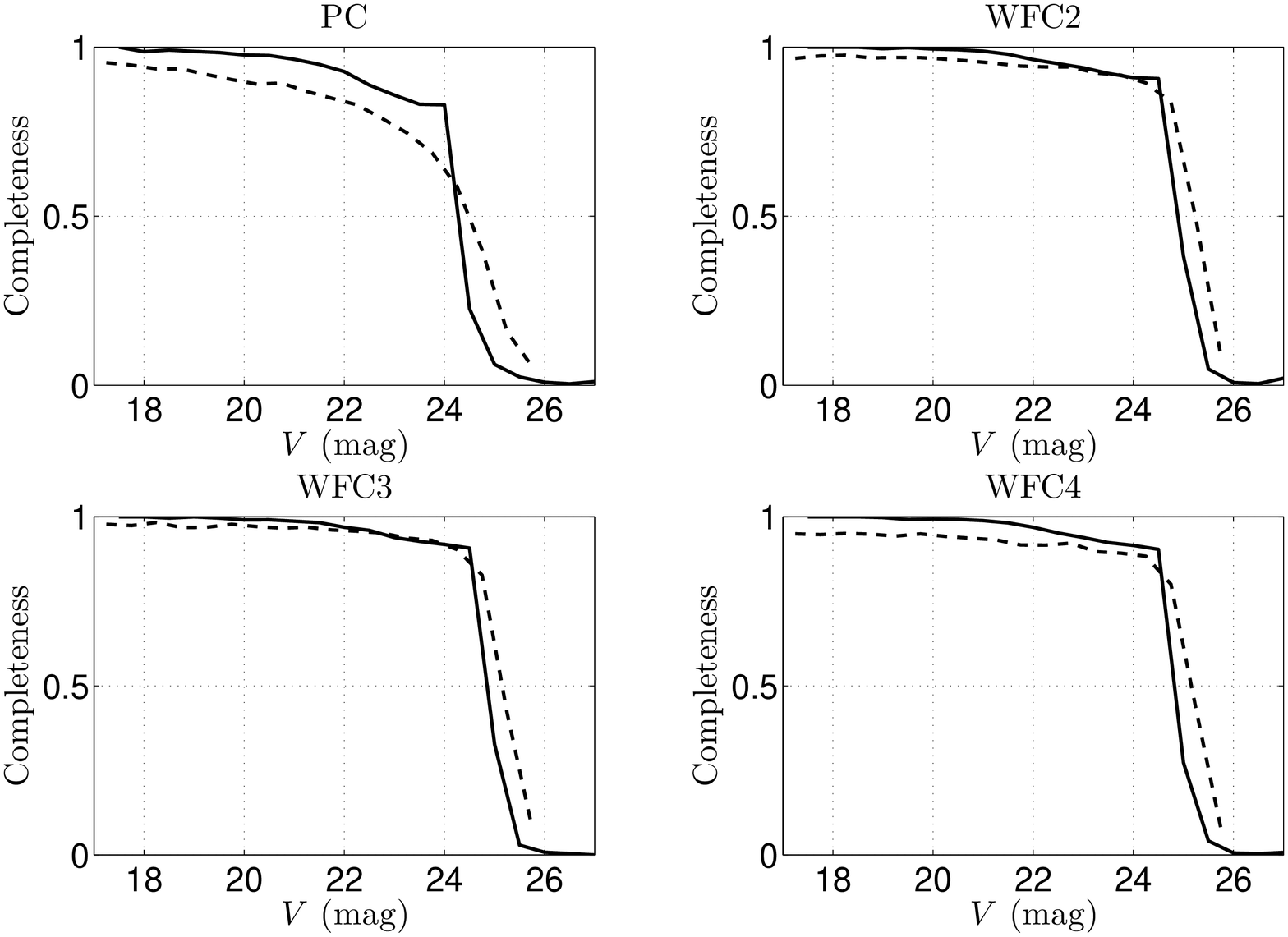}
 \caption{Completeness curves for the NGC 1805 data for different
   WFPC2 chips. Dashed, solid lines: Results from the real and
   catalogue-based \citep{Hu11} ASs tests, respectively.}
  \label{F18a.eps}
\end{figure}

\begin{figure}
 \includegraphics[width=90mm]{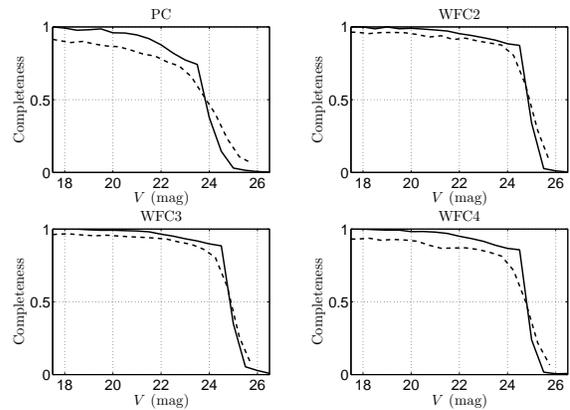}
 \caption{As Fig. \ref{F18a.eps}, but for NGC 1818.}
  \label{F18b.eps}
\end{figure}

We will now consider the origin of the main differences between both
methods. \cite{Hu11} mainly considered two factors that will affect
the incompleteness levels of a data set:
\begin{enumerate}
  \item For bright stars, incompleteness is mainly caused by
    star--star blends;
  \item For faint stars, incompleteness is mainly owing to stars
    remaining hidden in the background noise or below the background
    level.
\end{enumerate}
\cite{Hu11} mention that stars may also blend with bad pixels, cosmic
rays and -- to a very small extent -- background galaxies. They state
that these effects render their completeness levels lower than the
real completeness level which one could determine on the basis of
actual AS tests. We concur with this statement and we also emphasize
that special attention must be paid to saturated stars. The latter
objects will dominate the central regions of star cluster
observations, where it transpires that genuine stars will likely also
blend with the overspill counts (`bleeding' across detector rows or
columns) caused by saturated pixels. This will further reduce the
actual completeness levels of one's data set in high-density regions
where saturated stars may be common.

To explore this effect, we plotted the spatial distribution on the PC
chip of all ASs that were added to the NGC 1818 data but which were
not recovered by our actual AS tests. We also found that a significant
fraction of our source detections in the master catalogue originating
from the PC chip had been identified by {\sc HSTphot} as `extended'
sources. We became suspicious of the higher-than-expected fraction of
such sources and also plotted their distribution across the PC
chip. It turns out that this distribution displays a clear pattern
which is associated with the distribution of the saturated sources:
they trace cross-like patterns on the chip.

In Fig. 21 we show the distribution of both the `extended' sources
(red dots) and the ASs which our routines failed to recover (blue
dots). Both distributions are clearly similar. This suggests that
saturation will lead to enhanced incompleteness levels over and above
those expected from crowding and a higher background level in such
areas. Since the effects of mass segregation will cause the more
massive and, hence, brighter stars to migrate to a cluster's centre,
saturation will most likely disproportionately affect the
highest-density cluster core region. Hence, consideration of only real
point sources as genuine stars and adoption of a minimum blending
distance alone cannot reproduce the actual levels of sampling
(in)completeness in the presence of saturated stars. This is why, for
our observations on the PC chip, the catalogue-based AS tests yield a
significantly higher completeness fraction than that suggested by our
actual AS tests. Although the low-density data on the WF3 chip also
include a number of saturated stars, their number is much smaller, so
that the results from both methods are in much better agreement with
one another.

\begin{figure}
 \includegraphics[width=90mm]{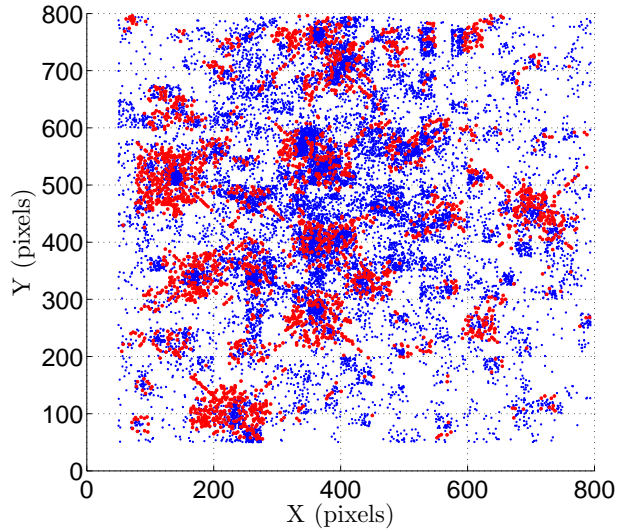}
 \caption{Spatial distributions of both the observed `extended'
   sources (red dots) and the actual ASs which were not recovered by
   our analysis routines. The cross-like patterns due to saturated
   stars are clearly discernible.}
  \label{F18c.eps}
\end{figure}

To further test if our conclusions are correct, we adapted the
incompleteness criteria of \cite{Hu11}. For stars brighter than the
detection limit, these authors assume that in a scenario where a faint
star blends with a brighter counterpart, the faint star will not be
recovered and only the brighter star is retained in the output
catalogue. We extended this criterion to all objects, including the
sources erroneously flagged as `extended'. No matter how bright the AS
is, once it blends with such `extended' sources, it will not be
recovered. In other words, even if the `extended' source is fainter
than the AS with which it is found to blend, the bright AS will not be
recovered, which hence reduces the data set's inferred completeness
levels.

We applied this new approach to the NGC 1818 data on the PC chip,
since there our catalogue-based approach most poorly reproduced the
completeness curve determined by the actual AS tests. Indeed, we found
that after implementation of this correction, the catalogue-based,
simulated completeness curves are much closer to the equivalent curves
based on the actual AS tests. However, we also noted a relatively
larger offset at the faint tail. This may have been caused by the
complicated LF on the PC chip due to the dynamical redistribution of
stars of different masses which leads to mass segregation. Figure 22
shows a comparison of the completeness curves based on the simulated
(catalogue-based) AS tests (blue dashed line), the actual AS tests
(black solid line) and the simulated AS tests after correction (blue
solid line), for the PC observations covering NGC 1818. In this paper
we use (and correct for) the (in)completeness levels inferred from the
actual AS tests.

\begin{figure}
 \includegraphics[width=90mm]{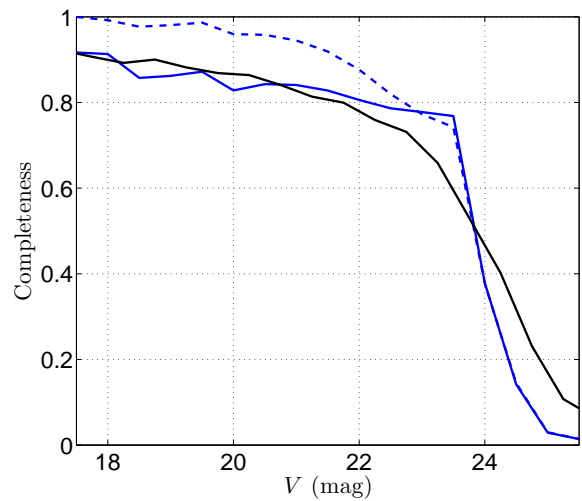}
 \caption{Completeness curves based on the simulated AS tests (blue
   dashed line), the actual AS tests (black solid line) and the
   simulated AS tests after correction (blue solid line; see text),
   applied to the NGC 1818 observations on the PC chip.}
  \label{F18d.eps}
\end{figure}

These results thus support our suggestion that saturated stars may
have a disproportionate effect on one's completeness levels. To prove
this beyond doubt, one would need to mask the pattern of saturated
stars and repeat the AS tests under different conditions and in areas
of different object densities. This is beyond the scope and remit of
the present paper, however. We will explore these issues in a future,
technical contribution.

Finally, we found that even though the simulated, catalogue-based AS
tests may fail to recover the full completeness characteristics in
crowded regions, our results regarding the occurrence of optical pairs
remain robust and unchanged. (The latter result was based on the
approach proposed by Hu et al. 2011.) This is so, because only stars
that blend with other stars will produce a `blending binary'. Stars
which are located close to or on the patterns caused by saturated
stars will either not be recovered or be identified as `extended'
objects. In either case, these blends are not included in our
analysis. The top panel of Fig. 23 compares the 3D blending fraction
of NGC 1805 obtained from simulated, catalogue-based AS tests (open
grid) with that resulting from the actual AS tests (filled grid). We
also show the total blending fraction as a function of radius over the
magnitude range of interest in this paper, $V \in [20.5, 22]$ mag
(bottom panel). Given that the typical uncertainties are $\lesssim 1$
per cent, both distributions are clearly very similar.

\begin{figure}
 \includegraphics[width=90mm]{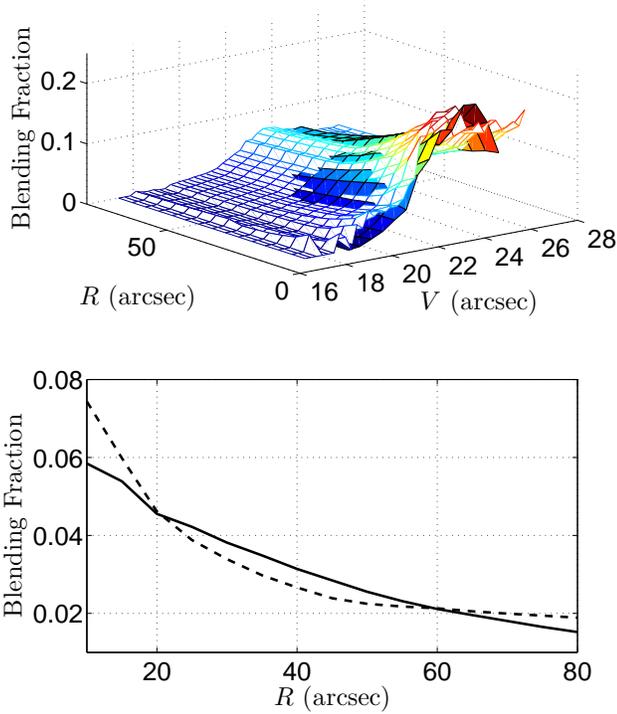}
 \caption{(top) 3D surface blending fraction as a function of radius
   and magnitude. Open, filled grids: Results from the simulated
   (catalogue-based) and actual AS tests, respectively. (bottom)
   Blending fraction for all stars with $V \in [20.5, 22]$ as a
   function of radius. Typical uncertainties are $\lesssim 1$ per
   cent.}
  \label{Blending.eps}
\end{figure}
 
\label{lastpage}
\end{document}